\documentclass[letterpaper,10pt]{article}

\usepackage[utf8]{inputenc}
\usepackage{amsmath}
\usepackage{amssymb}
\usepackage{amsfonts}
\usepackage{psfrag}
\usepackage{graphicx}
\usepackage{epsfig}

\usepackage{subfigure}

\usepackage[stable]{footmisc}

\title{Quasinormal frequencies of the dimensionally reduced BTZ black hole}

\author{ K.\ A.\ Guti\'errez-D\'avila$^{1}$, A.\ L\'opez-Ortega$^{1,*}$ \\
$^{1}$Departamento de F\'{\i}sica, 
Escuela Superior de F\'{\i}sica y Matem\'aticas, \\
Instituto Polit\'ecnico Nacional, \\
Unidad Profesional Adolfo L\'opez Mateos. Edificio 9. \\
Ciudad de M\'exico, M\'exico. \\
C.\ P.\ 07738 \\
Correspondence$^{*}$: A.\ L\'opez-Ortega\\
alopezo@ipn.mx
}

\begin{document}

\maketitle

\begin{abstract}

 We calculate numerically the quasinormal frequencies of the Klein-Gordon and Dirac fields moving in the two-dimensional dimensionally reduced BTZ black hole. Our work extends results previously published on the damped oscillations of this black hole.  First, we compute the quasinormal frequencies of the minimally coupled Klein-Gordon field for a range of the dimensionally reduced BTZ black hole physical parameters that is not previously analyzed. Furthermore we determine, for the first time, the quasinormal frequencies of the Dirac field propagating in the dimensionally reduced BTZ black hole. For the Dirac field we use the Horowitz-Hubeny method and the asymptotic iteration method, while for the Klein-Gordon field  the extension of the previous results is based on the asymptotic iteration method. Finally we discuss our main results.

KEYWORDS:  Quasinormal frequencies, Klein-Gordon and Dirac fields,  2D dimensionally reduced BTZ black hole, Horowitz-Hubeny method, Asymptotic iteration method

\end{abstract}

\section{Introduction}
\label{s: Introduction}

In the intermediate stage a perturbed black hole relaxes towards equilibrium performing a set of characteristic oscillations called quasinormal modes (QNMs in what follows). These modes are damped and the boundary condition that satisfy  near the event horizon is that the perturbation is purely ingoing \cite{Kokkotas:1999bd}--\cite{Berti:2009kk}. At the asymptotic region the boundary condition that fulfill the QNMs depends on the asymptotic structure of the black hole. For example, for asymptotically anti-de Sitter black holes usually is imposed that the perturbation field goes to zero at the asymptotic region \cite{Chan:1996yk}-\cite{Burgess:1984ti}. The frequencies of the QNMs are complex, the usually known as  quasinormal frequencies (QNFs in what follows), and we notice that the real part determines the oscillation frequency, whereas the imaginary part governs the decay rate. It is well known that the QNFs depend on the parameters of the black hole, the properties of the perturbation field, and they do not depend on the initial conditions of the perturbation. Also, these damped modes provide useful information on several characteristics of the black holes, such as the classical stability and the way they interact with the environment. More recently, the QNFs of asymptotically anti-de Sitter black holes have found applications in the framework of the AdS/CFT correspondence, since they are useful to determine the relaxation times in quantum field theories \cite{Kokkotas:1999bd}--\cite{Berti:2009kk}, \cite{Horowitz:1999jd}, \cite{Birmingham:2001pj}. 

Many  two-dimensional gravity models have been studied and at the present time we know  many exact solutions that represent two-dimensional black holes (2D black holes in what follows). This fact is useful since the theories of two-dimensional gravity allow us to study in more detail some classical and quantum properties of the black holes \cite{Grumiller:2002nm}, \cite{Grumiller:2006rc}. For example, for classical fields moving in 2D black holes, their equations of motion simplify and this fact allows us to calculate exactly some properties as the values of their QNFs. Some examples in which it is possible to calculate exactly the QNFs of 2D black holes are given in Refs.\ \cite{Becar:2007hu}--\cite{Zelnikov:2008rg}, however there are some 2D black holes for which a numerical computation is necessary to determine their QNFs \cite{Cordero:2012je}, \cite{Hernandez-Velazquez:2021zoh}.

We notice that  several 2D black holes are obtained by making a dimensional reduction to higher dimensional solutions \cite{Grumiller:2002nm}, \cite{Grumiller:2006rc}. An example is the dimensional reduction of the BTZ black hole proposed by Achucarro and Ortiz \cite{Achucarro:1993fd}. From the static three-dimensional BTZ black hole they obtain the so-called Jackiw-Teitelboim black hole (JT black hole in what follows). Also, from the rotating three-dimensional BTZ black hole they find the usually known as dimensionally reduced BTZ black hole (DRBTZ black hole in what follows \cite{Bhattacharjee:2020nul}). The properties of these 2D black holes are  widely studied \cite{Grumiller:2002nm}, \cite{Grumiller:2006rc}, \cite{Cordero:2012je}, \cite{Bhattacharjee:2020nul}, \cite{Achucarro:1993fd}. For example, the QNFs of the minimally coupled Klein-Gordon field and the Dirac field propagating in the JT black hole are exactly calculated in Ref.\  \cite{Cordero:2012je}. In contrast to the three-dimensional BTZ black hole \cite{Birmingham:2001pj}, \cite{Cardoso:2001hn}, for the minimally coupled  Klein-Gordon field moving in the DRBTZ black hole, the solutions to its radial equation are not expanded in terms of hypergeometric functions and a numerical calculation is necessary to determine its QNFs. Thus in Ref.\   \cite{Cordero:2012je},  taking as a basis the Horowitz-Hubeny method \cite{Horowitz:1999jd}, they numerically calculate the QNFs of the minimally coupled Klein-Gordon field. Nevertheless, in order to the Horowitz-Hubeny numerical method converges,  a restriction on the value of the inner horizon radius was imposed in Ref.\  \cite{Cordero:2012je}.\footnote{In Ref.\ \cite{Cordero:2012je} are studied the same 2D black holes that in this paper, but the Jackiw-Teitelboim black hole is called  ``uncharged Achucarro-Ortiz black hole'', whereas the dimensionally reduced BTZ black hole is called ``charged Achucarro-Ortiz black hole''. Since the last name is similar to one previously used to denote a three-dimensional black hole related to the BTZ solution (see for example  \cite{Pourhassan:2018chj} and references cited therein), in this paper, to avoid confusion, we do not employ the names given to these 2D black holes in Ref.\ \cite{Cordero:2012je}.   } 

In this work, our main objective is to extend the results of Ref.\ \cite{Cordero:2012je}. We make this extension in two ways. First, we calculate the QNFs of the minimally coupled Klein-Gordon field for a larger range than in Ref.\ \cite{Cordero:2012je} of the DRBTZ black hole physical parameters. Also our work is an extension of some results previously published in Ref.\ \cite{Bhattacharjee:2020nul} on the QNFs of the Klein-Gordon field propagating on the DRBTZ black hole  and our results extend those of Ref.\ \cite{Cadoni:2021qfn} on the QNFs of a scalar field propagating on the JT black hole. In Refs.\ \cite{Bhattacharjee:2020nul}, \cite{Cadoni:2021qfn} they use a modified Klein-Gordon equation. In contrast to \cite{Bhattacharjee:2020nul}, \cite{Cadoni:2021qfn} and in a similar way to  \cite{Cordero:2012je}, in what follows we calculate the QNFs of  a minimally coupled Klein-Gordon field to study the effect on its propagation of the DRBTZ black hole geometry. Furthermore we comment that the QNFs of the minimally coupled Klein-Gordon field are commonly studied in several backgrounds \cite{Kokkotas:1999bd}--\cite{Berti:2009kk}. Second, in the DRBTZ black hole  we numerically compute the QNFs of the Dirac field that were not calculated in the previous references. 

Our results allow us to study the following: i) The classical stability under perturbations of the 2D DRBTZ black hole. ii) The response of the DRBTZ black hole to Klein-Gordon and Dirac perturbations. Furthermore, using our results we compare the response of the studied black holes to the minimally coupled Klein-Gordon field and to the generalized Klein-Gordon field, previously examined in Refs.\ \cite{Bhattacharjee:2020nul}, \cite{Cadoni:2021qfn}. In a similar way, our results allow us to compare the response of the BTZ black hole and the dimensional reduction proposed by Achucarro-Ortiz. iii) The dependence of the QNFs on the physical parameters of the black hole and the field. Therefore we extend the results of Ref.\ \cite{Cordero:2012je} to analyze whether the dependence on the black hole physical parameters is of the same form for the values that are not previously explored  and we calculate the spectrum of QNFs for the Dirac field. For these reasons we think that it is useful to explore the spectrum of QNFs of the 2D DRBTZ black hole.

The rest of this paper is organized as follows. In the following section we give the relevant properties for our work of the DRBTZ black hole. In Sect.\ \ref{s: QNF-Klein-Gordon} we extend the previous results on the QNFs of the minimally coupled Klein-Gordon field propagating  in the DRBTZ black hole. In contrast to Ref.\ \cite{Cordero:2012je} that used the Horowitz-Hubeny numerical method \cite{Horowitz:1999jd}, in this section our computations are based on the Asymptotic Iteration Method (AIM in what follows) \cite{Ciftci Hall and Saad 2003}--\cite{Cho:2011sf}. This method allows us to analyze a larger range  of the DRBTZ black hole physical parameters. In Sect.\ \ref{s: QNF Dirac} we present our numerical results on the QNFs of the Dirac field moving in the DRBTZ black hole. For this field, in the computation we use the AIM and Horowitz-Hubeny method, but the first method allows us to explore a larger range of the DRBTZ black hole  physical parameters.  We discuss our main results in Sect.\ \ref{s: Discussion}. We describe the main characteristics of the AIM in Appendix \ref{a: aim method} and finally in Appendix \ref{a: HH Dirac}, for the Dirac field propagating in the DRBTZ black hole we expound the mathematical steps to transform its radial equations  into a convenient form to use the Horowitz-Hubeny method in the calculation of its QNFs.

\section{Dimensionally reduced BTZ black hole}
\label{s: cao black hole}

The DRBTZ black hole is a dimensional reduction of the three-dimensional rotating BTZ black hole. This 2D back hole is a solution to the equations of motion for the action \cite{Achucarro:1993fd} 
\begin{equation}
 S = \int d^2 x \sqrt{-h} \phi \left( \mathcal{R} + 2 \Lambda - \frac{J}{2 \phi^4} \right),
\end{equation} 
where $h$ is the determinant of the metric, $\mathcal{R}$ is the scalar curvature, $\Lambda=1/l^2$ determines the anti-de Sitter radius $l$, $J$ is the angular momentum of the BTZ black hole and $\phi$ is the dilaton field. The line element of the 2D DRBTZ black hole is equal to 	
\begin{equation}\label{e: metric CAO}
ds^2= f(r) dt^2 - \frac{dr^2}{f(r)},
\end{equation}
where
\begin{equation}\label{e: f CAO}
 f(r) = \frac{r^2}{l^2} - M + \frac{J^2}{4r^2},
\end{equation}
and the dilaton field is equal to $\phi=r$. This black hole has two horizons located at
\begin{equation}\label{e: horizons CAO}
r_{\pm}^2=\frac{l^2}{2} \Big( M \pm \sqrt{M^2 -\frac{J^2}{l^2}} \Big),
\end{equation}
with $r_+$ being  the radius of the event horizon and $r_-$ being the radius of the inner horizon. Taking into account that 
\begin{equation}\label{e: M J CAO}
M=\frac{r_+^2+r_-^2}{l^2}, \qquad \qquad \qquad J=\frac{2r_+ r_-}{l},
\end{equation}
we find that the function (\ref{e: f CAO}) can be written as
\begin{equation}\label{eq:f(r)}
f(r)=\frac{(r^2-r_+^2)(r^2-r_-^2)}{l^2r^2}.
\end{equation}
In what follows we take units where $l=1$ and we assume that the horizons radii satisfy $r_+ > r_-$. It is convenient to notice that the line element of the JT black hole is obtained from (\ref{e: metric CAO}) and (\ref{e: f CAO}) by taking $J=0$ or equivalently  $r_-=0$.

\section{QNFs of the Klein-Gordon field}
\label{s: QNF-Klein-Gordon}

Using the Horowitz-Hubeny method, in Ref.\ \cite{Cordero:2012je} are calculated the QNFs of the minimally coupled Klein-Gordon field propagating in the 2D DRBTZ black hole. Nevertheless, in the previous reference, in order to the Horowitz-Hubeny method converges, it is imposed the restriction $r_+ > 2 r_-$ on the values of the horizons radii. Therefore only for values of the horizons radii fulfilling the previous condition, the QNFs of the Klein-Gordon field are previously calculated.  For the 2D DRBTZ black hole, by using the AIM \cite{Ciftci Hall and Saad 2003}--\cite{Cho:2011sf}, we can determine the QNFs of the Klein-Gordon field for a larger range of values  for the inner horizon radius not satisfying the previous condition that is imposed in the implementation of the Horowitz-Hubeny method of Ref.\ \cite{Cordero:2012je}. 

In an analogous way to Ref.\ \cite{Cordero:2012je} we define the QNMs of the DRBTZ black hole as the solutions to the equations of motion that satisfy the boundary conditions:

a) They are purely ingoing near the event horizon.

b) They go to zero at the asymptotic region ($r \to \infty$).

\begin{figure}[ht]
\begin{center}
\includegraphics[scale=.6,clip=true]{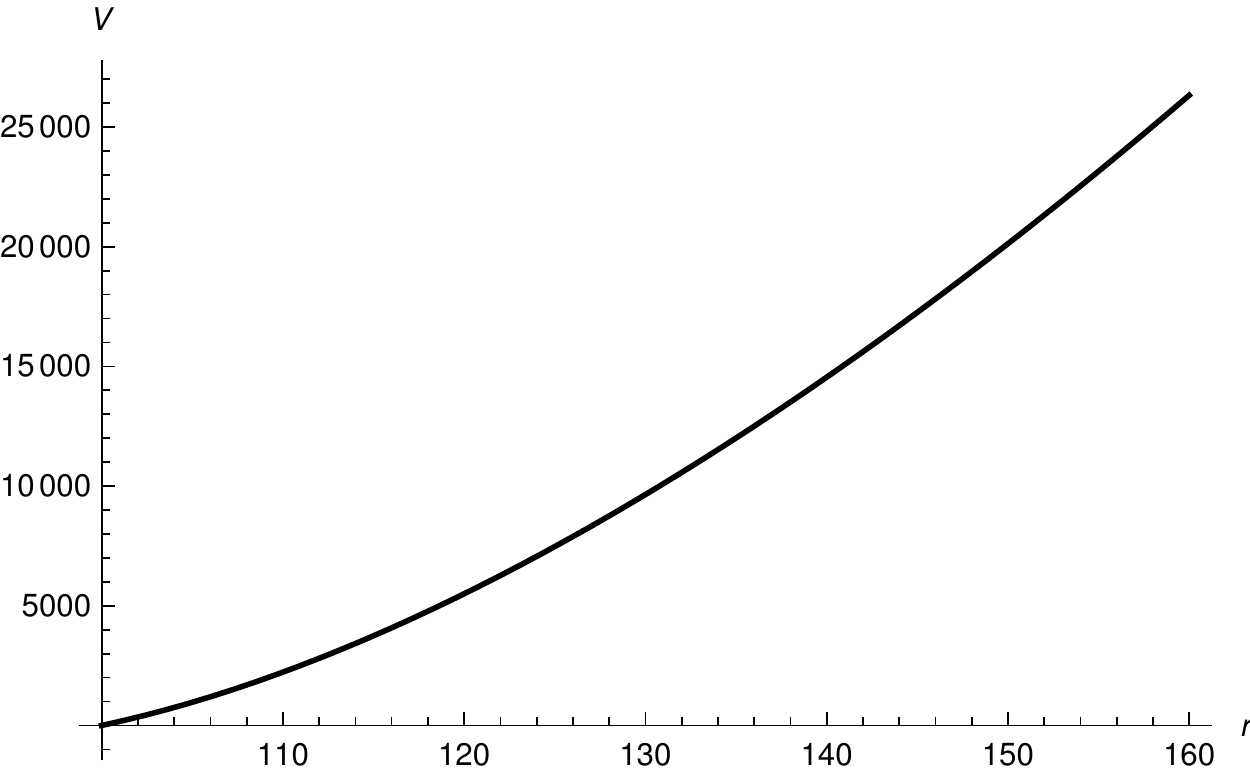}
\caption{We show the effective potential  (\ref{e: efective potential KG}) for the Klein-Gordon field. We take $m=3/2$, $r_+ = 100$ and $r_-=80$ ($M=16400$ and $J=16000$). \label{f: kg potential}} 
\end{center}
\end{figure}

As shown in Ref.\ \cite{LopezOrtega:2011sc}, in a two dimensional spacetime whose line element takes the form
\begin{equation}\label{e: metric general}
ds^2=f(r)dt^2 - \frac{dr^2}{g(r)},
\end{equation}
where the functions $f$ and $g$ depend only on the $r$ coordinate, the minimally coupled Klein-Gordon equation 
\begin{equation}\label{e:  KG eq}
 (\Box + m^2) \Phi = 0, 
\end{equation}
simplifies to the Schr\"odinger type equation
\begin{equation}\label{e: Schrodinger type KG}
\frac{d^2 R}{dr_*^2}+(\omega ^2 -V)R=0,
\end{equation}
when we take $\Phi = R(r) \textrm{exp}{(-i\omega t)}$. In the previous equation $r_*$ denotes the tortoise coordinate and in the metric (\ref{e: metric general}) is defined by
\begin{equation} \label{e: tortoise coordinate}
 r_*=\int \frac{dr}{\sqrt{f g}}.
\end{equation}
The effective potential $V$ is equal to  \cite{LopezOrtega:2011sc}
\begin{equation} \label{e: efective potential KG}
 V=m^2 f.
\end{equation} 
We notice that for $m=0$ the effective potential (\ref{e: efective potential KG}) goes to zero. Hence, for the minimally coupled Klein-Gordon field in the massless limit the solutions of the  radial equation (\ref{e: Schrodinger type KG}) are sinusoidal functions. As a consequence, in what follows we consider that the mass of the Klein-Gordon field is different from zero and in Fig.\ \ref{f: kg potential} we show the effective potential (\ref{e: efective potential KG}) for $m=3/2$, $r_+ = 100$ and $r_-=80$ ($M=16400$ and $J=16000$). We notice that the effective potential goes to zero at the horizon of the black hole and diverges as $r \to \infty$.

In the 2D DRBTZ black hole, Eq.\ (\ref{e: Schrodinger type KG}) takes the form \cite{Cordero:2012je}
\begin{equation}\label{e: aim kg radial}
 x(1-x) \frac{d^2R}{dx^2}+ \Big( 1-x-\frac{r_+^2-r_-^2}{2}\frac{x}{r_+^2-xr_-^2} \Big) \frac{dR}{dx} + \Big( \frac{\alpha}{x} + \gamma + \frac{\beta}{1-x} \Big) R = 0,
\end{equation}
where 
\begin{equation}
 \alpha = \frac{\omega ^2 r_+^2}{4(r_+^2-r_-^2)^2}, \quad \gamma = - \frac{\omega ^2 r_-^2}{4(r_+^2-r_-^2)^2}, \quad \beta = - \frac{m^2}{4},
\end{equation}
and the variable $x$ is defined by
\begin{equation}\label{e:aim kg new var}
 x=\frac{r^2-r_+^2}{r^2-r_-^2}. 
\end{equation}

To satisfy the boundary conditions a) and b) of the QNMs for the 2D DRBTZ black hole, we take the radial function $R$ as
\begin{equation}\label{e:aim kg solucion general}
 R = x^{-i \alpha ^{1/2}} (1-x)^{\frac{1}{4}(1 + \sqrt{1+4m^2})} \chi (x),
\end{equation}
to find that the function $\chi$ must be a solution of the differential equation 
\begin{equation}\label{e: aim kg aim radial}
 \frac{d^2 \chi}{dx^2} =\lambda _0(x) \frac{d \chi}{dx} + s_0 (x) \chi,
\end{equation}
where
\begin{align}\label{e:aim kg lambda0}
 \lambda _0 (x) &= \frac{2B+Cr_-^2/r_+^2}{1-x} + \frac{C(r_-^2/r_+^2)^2x}{(1-x)(1-xr_-^2/r_+^2)} - \frac{2A+1}{x},  \\
  s_0 (x) &= \frac{ACr_-^2/r_+^2}{x(1-x)(1-xr_-^2/r_+^2)} + \frac{B(r_-^2/r_+^2)}{2(1-x)^2} - \frac{BC(r_-^2/r_+^2)^2x}{(1-x)^2(1-xr_-^2/r_+^2)} \nonumber \\
 &- \frac{\alpha + \gamma + \beta -B(2A+1)}{x(1-x)}, \nonumber
\end{align}
with the quantities  $A$, $B$, and $C$ defined by
\begin{equation}\label{e: aim kg ABC}
 A=-i\alpha ^{1/2}, \qquad B=\frac{1}{4}(1+ \sqrt{1+4m^2}), \qquad C=\frac{r_+^2-r_-^2}{2r_-^2}.
\end{equation}
We notice that Eq.\ (\ref{e: aim kg aim radial}) has the mathematical form required by the AIM \cite{Ciftci Hall and Saad 2003}--\cite{Cho:2011sf} (see also Appendix \ref{a: aim method}). Hence, in the following section, taking as a basis Eq.\ (\ref{e: aim kg aim radial}) and using the AIM, we numerically calculate the QNFs of the minimally coupled Klein-Gordon field moving in the DRBTZ black hole.

\subsection{Numerical results}
\label{ss: numerical results KG}

In Ref.\ \cite{Cordero:2012je} are published the QNFs of the minimally coupled Klein-Gordon field propagating in the DRBTZ black hole. In order to the Horowitz-Hubeny method converges, in the previous reference the restriction $r_+ > 2 r_-$ was imposed. Using the AIM we reproduce the results presented in the previous reference for $r_+ > 2 r_-$ and in what follows we mainly show the QNFs of the Klein-Gordon field moving in the  DRBTZ black hole whose horizons radii satisfy $ r_- > r_+ / 2 $.\footnote{We point out that in this work the radii of the horizons must always fulfill $r_+ >  r_-$.}  We note that for the physical configurations for which the two methods work, the produced numerical results are in agreement.

We recall that in the JT black hole an analytical expression for  the QNFs  of the minimally coupled Klein-Gordon field     is known \cite{Cordero:2012je}
\begin{equation} \label{e: QNF KG UAO}
 \omega_{KG} =-i \kappa \left( n + \frac{1}{2}+\frac{1}{2} \sqrt{1+4 m^2} \right) ,
\end{equation} 
where $n$ is the mode number and $\kappa = r_+$ is the surface gravity of the JT black hole. We notice that the QNFs (\ref{e: QNF KG UAO}) are purely imaginary and we also comment that for the Klein-Gordon field propagating in the DRBTZ black hole, we numerically obtain purely imaginary QNFs (see also Refs.\ \cite{Cordero:2012je}, \cite{Bhattacharjee:2020nul}, \cite{Cadoni:2021qfn}). Therefore in our figures we plot the imaginary part of the QNFs and in what follows for the Klein-Gordon field we do not make comments on the real part of its QNFs.

\begin{figure}
  \centering
  \subfigure[$r_+=50$, $r_-=10$ ($M=2600$, $J=1000$)]{\includegraphics[scale=.4]{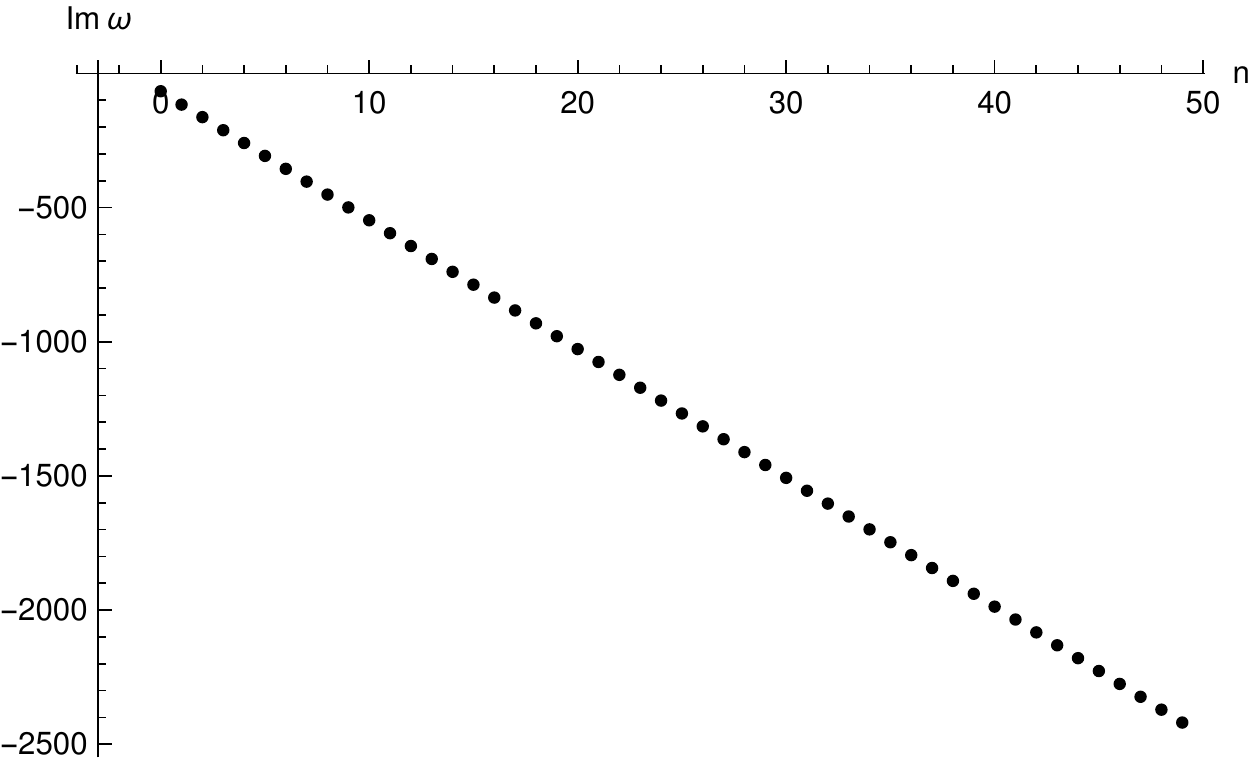}}
  \subfigure[$r_+=100$, $r_-=10$ ($M=10100$, $J=2000$)]{\includegraphics[scale=.4]{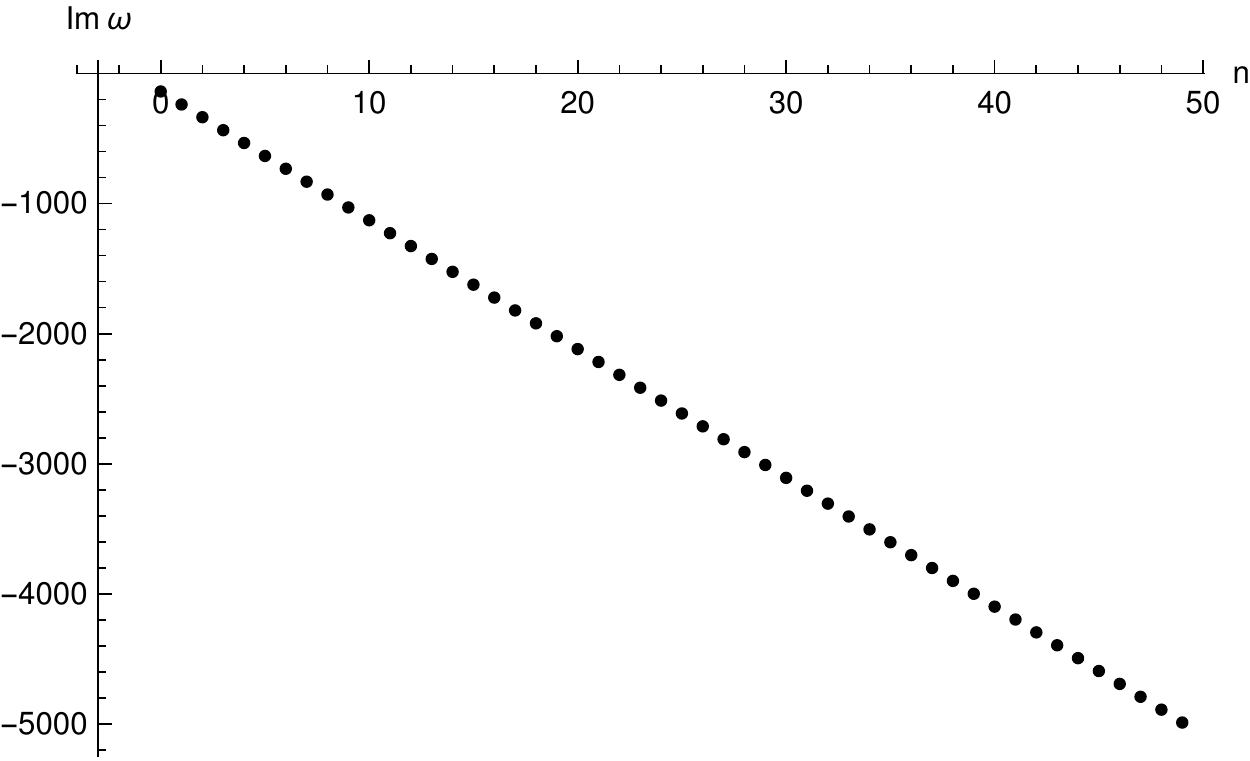}}
  \subfigure[$r_+=50$, $r_-=40$ ($M=4100$, $J=4000$)]{\includegraphics[scale=.4]{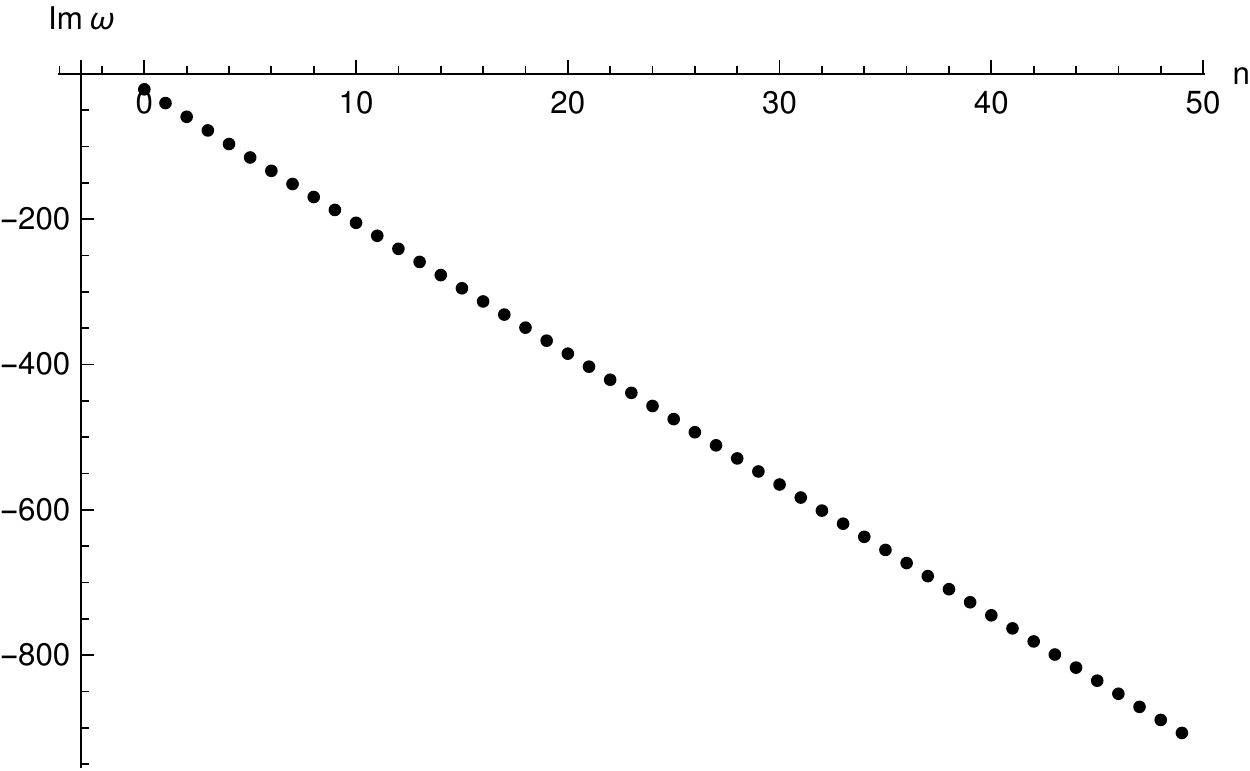}}
  \subfigure[$r_+=100$, $r_-=80$ ($M=16400$, $J=16000$)]{\includegraphics[scale=.4]{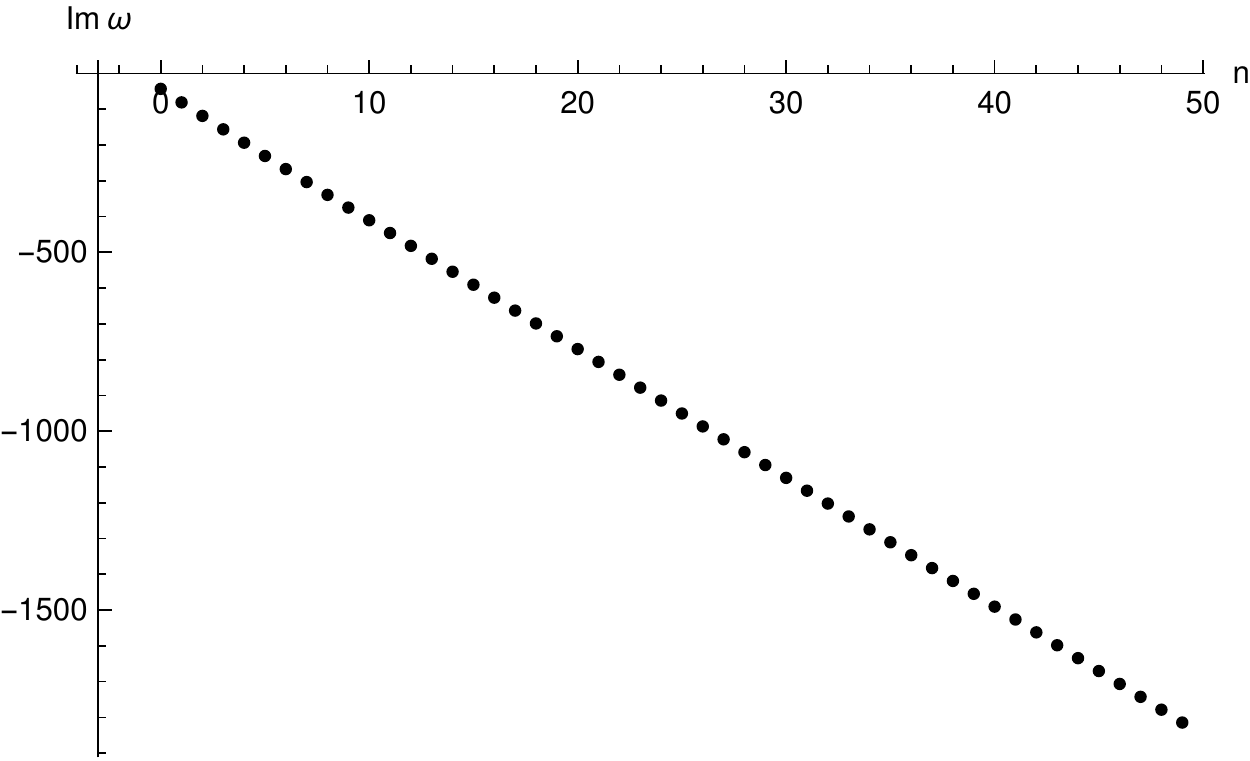}}
   \caption{We show the first fifty QNFs of the Klein-Gordon field for different values of the horizons. We take $m=3/4$.}\label{f: kg n variable}
\end{figure}

\begin{table}[h]
\centering
\begin{tabular}{cccc}
\hline
$r_+$ & $r_-$ & $\kappa$ & Linear fit  \\  \hline
50 & 10  & 48   &  $-67.222 - 48.001 n$  \\ 
 100 & 10 & 99  & $-138.715 - 99.000 n $ \\ 
 50 &  40 & 18  &  $-24.566 - 18.019 n $ \\ 
 100 &  80 & 36 & $-49.091 - 36.040n $ \\ \hline
\end{tabular}
\caption{For the  first fifty QNFs of the Klein-Gordon field we give the linear fits for the four examples of Fig.\ \ref{f: kg n variable}. We note that this figure shows the dependence of the QNFs on the mode number.}
\label{Tabla 1}
\end{table}

For the Klein-Gordon field, in Fig.\ \ref{f: kg n variable}  we plot the first fifty QNFs for four different combinations of $r_+$ and $r_-$.  For the presented examples, we see that the plots of Fig.\ \ref{f: kg n variable} show a linear relation between ${\mathbb{I}}{\mathrm{m}} (\omega)$ and $n$.  To analyze this behavior in Table \ref{Tabla 1} we  give the linear fits for the four graphs of Fig.\ \ref{f: kg n variable}. In Table \ref{Tabla 1}  we observe that for the linear fit the absolute value of the slope is almost identical with the value of the surface gravity, that for the DRBTZ black hole  is given by \cite{Achucarro:1993fd}
\begin{equation}\label{e: surface gravity}
 \kappa=\frac{r_+^2-r_-^2}{r_+} .
\end{equation} 
This behavior is similar to the one we deduce from the exact expression for the QNFs of the JT black hole, but for the DRBTZ black hole the slope is not exactly its surface gravity.

\begin{figure}[ht]
\begin{center}
\includegraphics[scale=.9,clip=true]{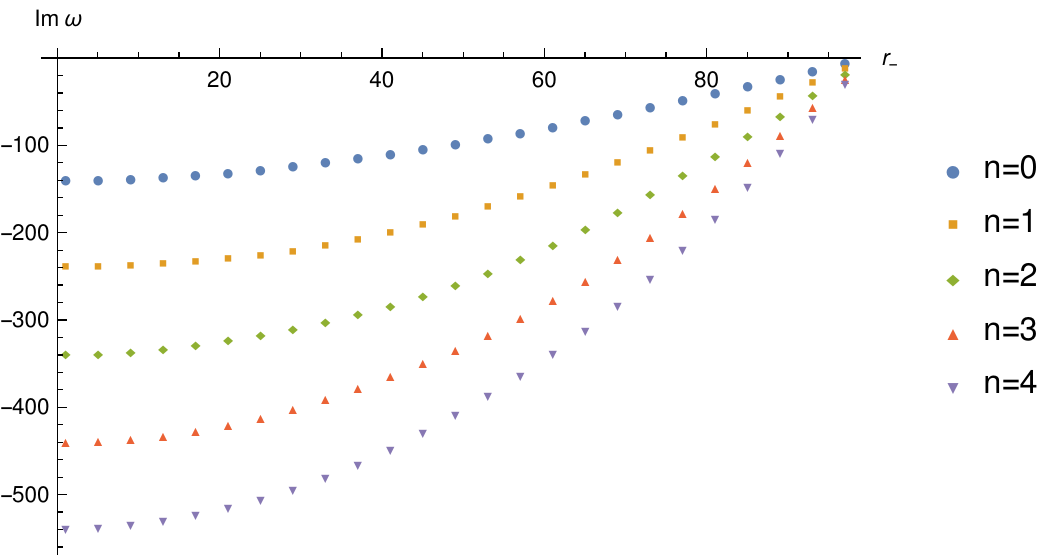}
\caption{For the Klein-Gordon field of mass $m=3/4$ we show the dependence on the inner horizon radius of the imaginary part for the first five QNFs. We take $r_+=100$.  \label{f: kg rmenos variable}} 
\end{center}
\end{figure}

\begin{table}[ht]
\centering
\begin{tabular}{cc}
\hline
$n$ & Quadratic fit  \\  \hline
0 & $-142.933 + 0.366 r_-  + 0.011 r_-^2$   \\ 
 1 & $ -241.953 + 0.039 r_- + 0.024 r_-^2$  \\ 
 2 & $-338.972 - 0.089 r_- + 0.035 r_-^2 $  \\ 
 3 & $-439.832 - 0.044 r_- + 0.045 r_-^2 $  \\ 
 4 & $ -539.519 - 0.050 r_- + 0.055 r_-^2$   \\  \hline
\end{tabular}
\caption{For the Klein-Gordon field we give the quadratic fits for the QNFs of the first five modes of Fig.\ \ref{f: kg rmenos variable}. We observe that this figure shows the dependence of the QNFs on the inner horizon radius.}
\label{Tabla 2}
\end{table}

\begin{table}[ht]
\centering
\begin{tabular}{cccc}
\hline
$n$ & $\omega_{JT}^a$  & $\omega_{JT}^q$ & $\mathcal{E}$  \\  \hline
0 &$-140.139 i$ & $-142.933 i$  &   1.994 \%\\ 
 1 & $-240.139 i$& $ -241.953i $ & 0.755 \% \\ 
 2 &$ -340.139 i$& $-338.972 i$  &  0.343 \% \\ 
 3 & $-440.139 i$& $-439.832 i $ & 0.070 \% \\ 
 4 & $-540.139 i$& $ -539.519 i$ &  0.115 \% \\  \hline
\end{tabular}
\caption{Based on the quadratic fits of Table \ref{Tabla 2} we give the predicted values for the QNFs of the Klein-Gordon field moving in the JT black hole ($J=r_-=0$) whose QNFs are known analytically \cite{Cordero:2012je}. We also show the relative errors.}
\label{Tabla 3}
\end{table}

For the Klein-Gordon field of mass $m=3/4$, in Fig.\ \ref{f: kg rmenos variable}  we show how the QNFs depend on the inner horizon radius for a DRBTZ black hole of event horizon radius $r_+=100$. In this figure we observe that the absolute value of the imaginary part of the QNFs decreases as the inner horizon radius increases, that is, the decay time increases as the inner horizon radius increases. The decay time $\tau$ is equal to
\begin{equation}
 \tau=\frac{1}{{\lvert \mathbb{I}}{\mathrm{m}} (\omega_0) \rvert} ,
\end{equation} 
where $ {\mathbb{I}}{\mathrm{m}} (\omega_0)$   is the imaginary part of the fundamental mode.  Furthermore we observe that for the first five modes the damping decreases as the inner horizon radius increases.\footnote{ Since for constant event horizon radius, the quantities $J$ and $r_-$ are proportional (see the expression (\ref{e: M J CAO})), for the graph ${\mathbb{I}}{\mathrm{m}} (\omega)$ vs $J$ we get a similar behavior to that shown in Fig.\  \ref{f: kg rmenos variable}.}

We also see that the variation of the imaginary part of the QNFs with the inner horizon radius can be described by a quadratic relation. For the first five QNFs of Fig.\  \ref{f: kg rmenos variable}  we give the quadratic fits in Table \ref{Tabla 2}. For the QNFs of the Klein-Gordon field propagating in the JT black hole, in Table \ref{Tabla 3} we show the  predicted values when we take as a basis the quadratic fits of Table \ref{Tabla 2} and compare with the values that produces the exact expression (\ref{e: QNF KG UAO}). We also give the relative error of the predicted values for the QNFs of the JT black hole. The relative error is defined by
\begin{equation} \label{e: relative error}
 \mathcal{E}=\left\lvert  \frac{\omega_{JT}^a-\omega_{JT}^q}{\omega_{JT}^a} \right\rvert \times 100\%,
\end{equation} 
where $\omega_{JT}^a$is the value that produces the analytical expression (\ref{e: QNF KG UAO}) and $\omega_{JT}^q$ is the predicted value  by the quadratic fit of Table \ref{Tabla 2}. We see that the quadratic fits of Table \ref{Tabla 2} generate accurate values for the QNFs of the JT black hole.

\begin{figure}[ht]
\begin{center}
\includegraphics[scale=.9,clip=true]{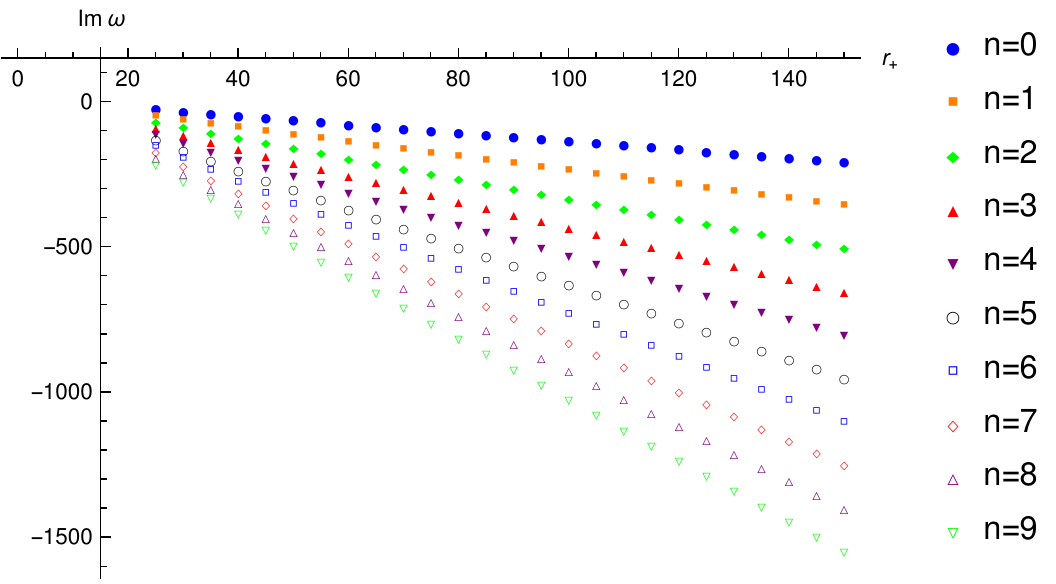}
\caption{We show the dependence on the event horizon radius of the imaginary part of the first ten QNFs for the Klein-Gordon field of mass $m=3/4$. We take $r_-=10$.  \label{f: kg rmas variable}} 
\end{center}
\end{figure}

\begin{table}[ht]
\centering
\begin{tabular}{cc}
\hline
$n$ & Linear fit  \\  \hline
0 &  $6.053 - 1.439 r_+$ \\ 
 1 &  $6.951 - 2.447 r_+$ \\ 
 2 &  $11.409 - 3.470 r_+$ \\ 
 3 & $14.316 - 4.494 r_+ $\\ 
 4 & $ 17.460 - 5.508 r_+ $\\ 
 5 & $ 20.979 - 6.536 r_+$ \\  
 6 & $ 23.926 - 7.549 r_+ $ \\  
 7 &  $ 27.514 - 8.576 r_+ $\\  
 8 &  $30.430 - 9.590 r_+ $ \\  
 9 &  $33.981 - 10.615 r_+ $ \\  \hline
\end{tabular}
\caption{For the Klein-Gordon field we display the linear fits for the QNFs of the first ten modes of Fig.\ \ref{f: kg rmas variable}. We notice that this figure shows the dependence on the event horizon radius of the QNFs.}
\label{Tabla 5}
\end{table}

For the DRBTZ black hole of inner horizon  radius $r_-=10$, in Fig.\ \ref{f: kg rmas variable} we show how the first ten QNFs depend on the value of the event horizon radius $r_+$. We observe that the imaginary part of the QNFs depends on  the value of the event horizon radius in a linear way. Furthermore in Table \ref{Tabla 5} we write the linear fits for the first ten QNFs presented in Fig.\ \ref{f: kg rmas variable}. From the expressions of Table \ref{Tabla 5} we see that the absolute value of the slope increases with the mode number and the value of the slope is approximately equal to $-(n+ 3/2)$, but we notice that for the first four QNFs the absolute value of the slope is less than $n+ 3/2$, whereas for the last five modes the absolute value of the slope is greater than $n+ 3/2$.  For $n=4$ the absolute value of the slope is almost equal to $n+ 3/2$.  That is, the slope changes as the mode number varies. This linear behavior is similar to the one we deduce from the exact QNFs of the JT black hole, but for the values of the parameters for our example, the absolute value of the slope for the JT is $n+ 1/2+\sqrt{13}/4 \approx n + 1.401$ for all the QNFs (see the formula (\ref{e: QNF KG UAO})). Thus for the DRBTZ black hole the absolute value of the slope for the plot ${\mathbb{I}}{\mathrm{m}} (\omega)$ vs $r_+$ is greater than the one for the JT black hole and the slope depends on the mode number. Furthermore, from Fig.\ \ref{f: kg rmas variable} we get that for fixed inner horizon radius, the decay time decreases as the event horizon radius increases. Also, for the first ten QNMs the damping increases as the event horizon radius increases.

\begin{figure}[ht]
\begin{center}
\includegraphics[scale=.9,clip=true]{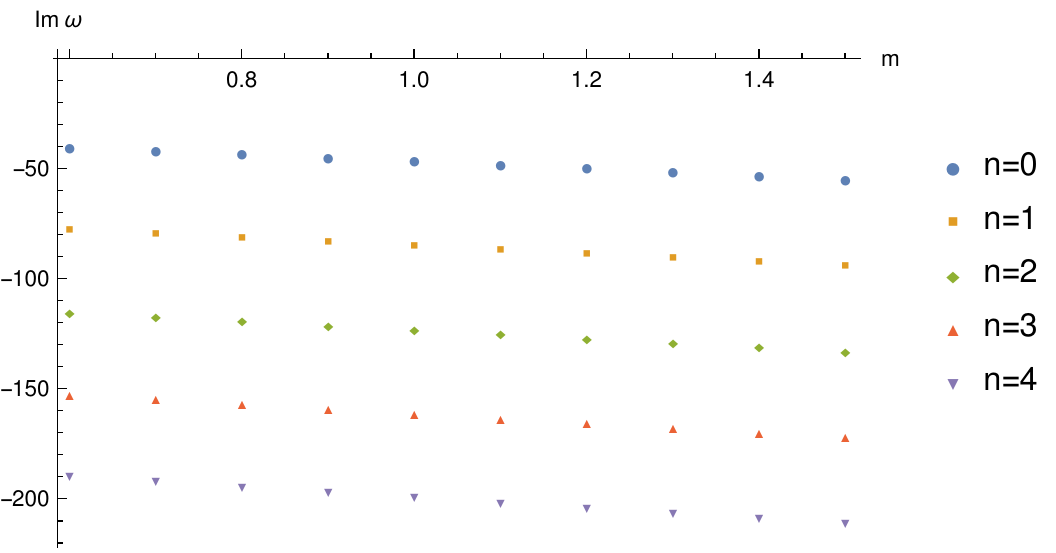}
\caption{We show the dependence on the field mass of the imaginary part of the QNFs for the Klein-Gordon field. We take $r_+=100$ and $r_-=80$ ($M=16400$, $J=16000$).  \label{f: kg mass variable}} 
\end{center}
\end{figure}

\begin{table}[ht]
\centering
\begin{tabular}{cc}
\hline
$n$ & Linear fit  \\  \hline
0 & $-30.93 - 16.08 m$ \\ 
 1 &  $-67.44 - 17.87 m$ \\ 
 2 &  $-103.83 - 19.67 m$ \\ 
 3 & $-140.01 - 21.55 m$ \\ 
 4 & $-175.88 - 23.56 m$ \\  \hline
\end{tabular}
\caption{For the Klein-Gordon field we display the linear fits for the QNFs of the first five modes of Fig.\ \ref{f: kg mass variable}. Notice that in this figure we show the dependence of the QNFs on the mass of the field. }
\label{Tabla 6}
\end{table}

In Fig.\ \ref{f: kg mass variable} we show how the imaginary part of the QNFs depends on the mass of the Klein-Gordon field moving in the DRBTZ black hole. We observe  that the relation between ${\mathbb{I}}{\mathrm{m}} (\omega)$ and $m$ is linear. In Table \ref{Tabla 6} we write the linear fits for the first five QNFs of Fig.\ \ref{f: kg mass variable}. From the data of this table we deduce that for the linear fits the absolute values of the slopes and the absolute values of the constant terms increase as the mode number increases. Also, from Fig.\ \ref{f: kg mass variable} we obtain that the decay time depends slightly on the mass, in such a way that the decay time decreases as the mass increases.

For the Klein-Gordon field, in each of Figures \ref{f: kg rmenos variable}, \ref{f: kg rmas variable}, \ref{f: kg mass variable}, we notice that  the analyzed  QNFs behave in a similar way as we change the corresponding physical parameter.

\section{QNFs of the Dirac field}
\label{s: QNF Dirac}

Following the usual method \cite{Parker-Toms}, we can show that in a bidimensional static spacetime of the form (\ref{e: metric general}) but written as 
\begin{equation}\label{e:NP Ecuacion Metrica}
 ds^2=P^2(r)dt^2-Q^2(r)dr^2,
\end{equation}
where the functions $P$ and $Q$ depend only on $r$, the Dirac equation 
\begin{equation}\label{e: Ecuacion Dirac}
 (i\gamma^{\mu} \nabla_{\mu} - m)\Psi = 0,
\end{equation}
simplifies to the coupled system of differential equations \cite{Hernandez-Velazquez:2021zoh}
\begin{align}\label{e: Ecuacion Dirac 2}
 \frac{1}{P}\partial_t \psi_2 - \frac{1}{Q}\partial_r \psi_2 - \frac{1}{2PQ} \frac{dP}{dr} \psi_2 & = -im \psi_1, \nonumber \\
  \frac{1}{P}\partial_t \psi_1 + \frac{1}{Q}\partial_r \psi_1 + \frac{1}{2PQ} \frac{dP}{dr} \psi_1 & = -im \psi_2,
\end{align}
when we use the null dyad
\begin{equation}\label{e: NP Base}
\hat{e}_1^{\; \mu} = \frac{1}{\sqrt{2}}\Big(\frac{1}{P},\frac{1}{Q}\Big), \qquad\hat{e}_2^{\; \mu}= \frac{1}{\sqrt{2}}\Big(\frac{1}{P},-\frac{1}{Q}\Big),
\end{equation}
and the representation of the gamma matrices 
\begin{equation}\label{e: NP Matrices Gamma}
 \gamma_{1} = \frac{1}{\sqrt{2}}\begin{pmatrix}
                                 0 & 2 \\
                                 0 & 0
                                \end{pmatrix}, 
\quad \gamma_{2} = \frac{1}{\sqrt{2}} \begin{pmatrix}
                                       0 & 0\\
                                       2 & 0
                                      \end{pmatrix}.
\end{equation}
We note that in Eq.\ (\ref{e: Ecuacion Dirac 2}) the quantities $\psi_1$ and $\psi_2$ are the components of the spinor $\Psi$, that is,
\begin{equation}\label{e: Espinor Dirac}
 \Psi = \begin{pmatrix}
         \psi_1 \\
         \psi_2
        \end{pmatrix}.          
\end{equation}

Taking a harmonic time dependence of the form
\begin{equation}\label{e: Dirac time dependence}
 \psi_1(t,r)=R_1(r)e^{-i\omega t}, \qquad \psi_2(t,r)=R_2(r)e^{-i\omega t},
\end{equation}
we find that the system of equations (\ref{e: Ecuacion Dirac 2}) reduces to
\begin{align}\label{e: Ecuacion Dirac harmonic}
 \frac{1}{Q}\frac{dR_2}{dr} + \frac{i \omega}{P}R_2 + \frac{1}{2PQ}\frac{dP}{dr}R_2 &=imR_1, \nonumber \\
  \frac{1}{Q}\frac{dR_1}{dr} - \frac{i \omega}{P}R_1 + \frac{1}{2PQ}\frac{dP}{dr}R_1 &=-imR_2,
\end{align}
from which we obtain decoupled equations for the radial functions $R_1$ and $R_2$. For 2D spacetimes that fulfill $(PQ)^2=1$, for example, the DRBTZ and JT black holes,  we find that these decoupled radial equations can be transformed into Schr\"odinger type equations of the form \cite{Hernandez-Velazquez:2021zoh}
\begin{equation}\label{e: schrodinger dirac}
\frac{dR_s}{dr_*^2}+(\omega^2-V_s)R_s=0, \qquad 
\end{equation}
where $s=1$ ($s=2$) for $R_1$ ($R_2$) and the effective potentials $V_s$ take the form
\begin{equation}\label{e: potenciales efectivos Dirac}
V_s=m^2f \mp \frac{i \omega}{2}\frac{df}{dr}-\frac{f}{4}\frac{d^2f}{dr^2}+\frac{1}{16}\Big(\frac{df}{dr}\Big)^2,
\end{equation}
with the function $f$ defined by $f=P^2$, in agreement with the expression (\ref{e: metric general}). In the formula (\ref{e: potenciales efectivos Dirac}) and in what follows the upper (lower) sign corresponds to $s=1$ ($s=2$).

\begin{figure}[ht]
\begin{center}
\includegraphics[scale=1.2,clip=true]{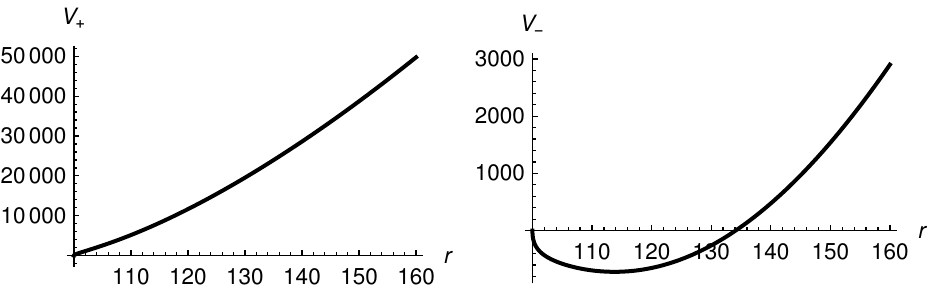}
\caption{We show the effective potentials $V_{\pm}$ for the Dirac field moving in the DRBTZ black hole.  We take $m=3/4$, $r_+=100$, $r_-=80$ ($M=16400$, $J=16000$).  \label{f: dirac potentials}} 
\end{center}
\end{figure}

For the Dirac field we notice that the effective potentials (\ref{e: potenciales efectivos Dirac}) are complex. Owing this fact we do not plot these effective potentials. We observe that they are different from the real effective potentials given in Ref.\ \cite{LopezOrtega:2011sc}, but in the last reference they used a different dyad and a different representation of the gamma matrices to write the Dirac equation. Furthermore, in Ref.\ \cite{LopezOrtega:2011sc} is shown that in a two-dimensional spacetime of the form (\ref{e:NP Ecuacion Metrica}), the Dirac equation simplifies to a pair of Schr\"odinger type equations with effective potentials 
\begin{equation}
 V_{\pm}= m^2 f \pm \left(\frac{m}{2}\right) \sqrt{f} \frac{d f}{dr},
\end{equation} 
when $(PQ)^2=1$ and $f=P^2$, as previously. We note that these effective potentials are real and for the DRBTZ black hole, we plot them in Fig.\ \ref{f: dirac potentials}. In a similar way to the effective potential for the Klein-Gordon field (\ref{e: efective potential KG}), the potentials $V_{\pm}$ of the Dirac field  go to zero at the horizon of the black hole and diverge as $r \to \infty$.

We have not been able to find exact solutions to the radial equations (\ref{e: schrodinger dirac}) in the DRBTZ black hole. Therefore, in what follows, we use numerical methods to find the QNFs of the Dirac field moving in the DRBTZ black hole. In Appendix \ref{a: HH Dirac} we expound the necessary steps to transform the differential equations (\ref{e: schrodinger dirac}) into an appropriate form to use the Horowitz-Hubeny method \cite{Horowitz:1999jd}, but as in Ref.\ \cite{Cordero:2012je}, the Horowitz-Hubeny method only works for DRBTZ black holes whose horizons radii satisfy $r_+ > 2 r_-$. Hence, in this section we transform Eqs.\ (\ref{e: schrodinger dirac}) into an appropriate mathematical form to use the AIM \cite{Ciftci Hall and Saad 2003}--\cite{Cho:2011sf} (see also Appendix \ref{a: aim method}), since  this numerical procedure works for a larger range of the DRBTZ black hole physical parameters. We first make the ansatz 
\begin{equation}\label{e: aim dirac ansatz}
 R_s=e^{-i(\omega \pm \frac{i\kappa}{2})r_*} r^{-(\frac{1}{2}+m)}\tilde{R}_s(r),
\end{equation}
to consider the boundary conditions a) and b) of the QNMs. In the previous equation, the factor $\textrm{exp}({-i(\omega \pm i\kappa/2)r_*})$  takes this form since the effective potentials (\ref{e: potenciales efectivos Dirac}) are different from zero at the event horizon of the DRBTZ black hole. For other fields, for example, the Klein-Gordon field, the effective potentials go to zero at the event horizon and as a consequence, near the event horizon we consider the boundary condition of the QNMs proposing a factor of the form  $\textrm{exp}(-i \omega r_*)$ \cite{Horowitz:1999jd}, \cite{Cordero:2012je}.

From Eqs.\ (\ref{e: schrodinger dirac}) we get that the functions $\tilde{R}_s$ are solutions to the differential equations 
\begin{equation}\label{e:cao dirac boundary}
 \frac{d^2\tilde{R}_s}{dr^2}+F_s(r) \frac{d\tilde{R}_s}{dr}+G_s(r) \tilde{R}_s=0,
\end{equation}
where
\begin{align}
 F_s(r)&=\frac{1}{f} \Big( \frac{df}{dr} - 2i\omega \pm \kappa \Big) - \frac{2(\frac{1}{2} + m)}{r}, \\
 G_s(r) &= \frac{\kappa ^2}{4f^2} \mp \frac{i \kappa \omega}{f^2} - \frac{V_s}{f^2} - \frac{(\frac{1}{2} + m)}{rf} \Big( \frac{df}{dr} - 2i \omega \pm \kappa \Big) + 
 \frac{(\frac{1}{2}+m)(\frac{3}{2}+m)}{r^2}. \nonumber 
\end{align}
Making the change of variable
\begin{equation} \label{e: new variable dirac}
 u=\frac{r-r_+}{r-r_-}
\end{equation} 
we find that Eqs.\ (\ref{e:cao dirac boundary}) transform into 
\begin{equation}\label{e: aim dirac final}
 \frac{d^2\tilde{R}_s}{du^2} = \lambda_{0,s} \frac{d\tilde{R}_s}{du} + s_{0,s}\tilde{R}_s 
\end{equation}
where
\begin{align}\label{e: lambda s aim dirac}
 \lambda_{0,s} & = - F_s(u) \frac{(r_+ - r_-)}{(1-u)^2} + \frac{2}{1-u} , \nonumber \\
 s_{0,s} & = - \frac{(r_+ - r_-)^2}{(1-u)^4} G_s(u).
\end{align}
We note that Eqs.\ (\ref{e: aim dirac final}) have the mathematical form that is appropriate to use the AIM (see Appendix \ref{a: aim method}) and taking them as a basis we can determine the QNFs of the Dirac field moving in the DRBTZ black hole. Furthermore, we observe that compared to the corresponding formula for the Klein-Gordon field,  the previous expressions for $\lambda_{0,s} $ and $s_{0,s}$ are more complicated and to determine the QNFs of the Dirac field the numerical computations will be more difficult. We comment that the QNFs of the Dirac field in the DRBTZ black hole are not previously calculated. To finish this section we notice that our numerical results show that the two effective potentials $V_s$ produce the same QNFs for the Dirac field.

\subsection{Numerical results}
\label{ss: numerical results Dirac}

In what follows we give the QNFs of the Dirac field propagating in the DRBTZ black hole. We mention that for the DRBTZ black hole the QNFs of this field were not previously computed in Refs.\ \cite{Cordero:2012je}, \cite{Bhattacharjee:2020nul}, \cite{Cadoni:2021qfn}. For $2r_- < r_+$ we can use the Horowitz-Hubeny method \cite{Horowitz:1999jd} and the AIM  \cite{Ciftci Hall and Saad 2003}--\cite{Cho:2011sf} to calculate the QNFs of the Dirac field moving in the DRBTZ black hole. For the horizons radii for which we can use the two numerical procedures, we find that both  methods  produce the same QNFs. For horizons radii satisfying $r_- > r _+ / 2 $ we use only the AIM, due to the Horowitz-Hubeny method does not converge for these values of the radii \cite{Cordero:2012je} (see also Appendix \ref{a: HH Dirac}). 

We also notice that the QNFs of the Dirac field with mass $m$ propagating in the JT black hole are calculated exactly in Ref.\ \cite{Cordero:2012je} and are equal to
\begin{equation} \label{e: dirac qnf uncharged}
 \omega_D =-i \kappa \left( m+\frac{1}{2}+n \right).
\end{equation} 
As in the formula (\ref{e: QNF KG UAO}) for the QNFs of the Klein-Gordon field, in the previous expression, $\kappa=r_+$ is the surface gravity of the JT black hole. In the expression (\ref{e: dirac qnf uncharged}) we notice that the QNFs of the Dirac field in the JT black hole are purely imaginary and for this field moving in the DRBTZ black hole we numerically obtain purely imaginary QNFs. Therefore, in what follows, for the Dirac field we only discuss the behavior of the imaginary part of its QNFs.

\begin{figure}
  \centering
  \subfigure[$r_+=50$, $r_-=10$ ($M=2600$, $J=1000$)]{\includegraphics[scale=.4]{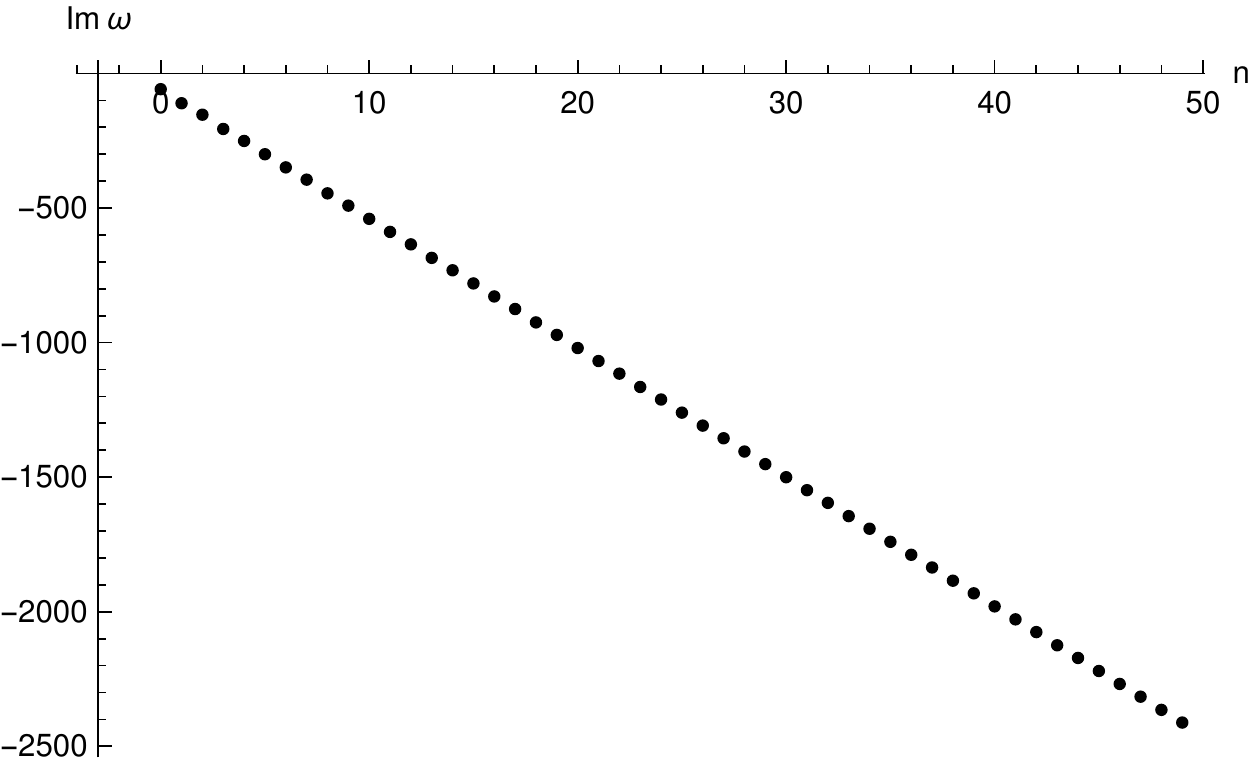}}
  \subfigure[$r_+=100$, $r_-=10$ ($M=10100$, $J=2000$)]{\includegraphics[scale=.4]{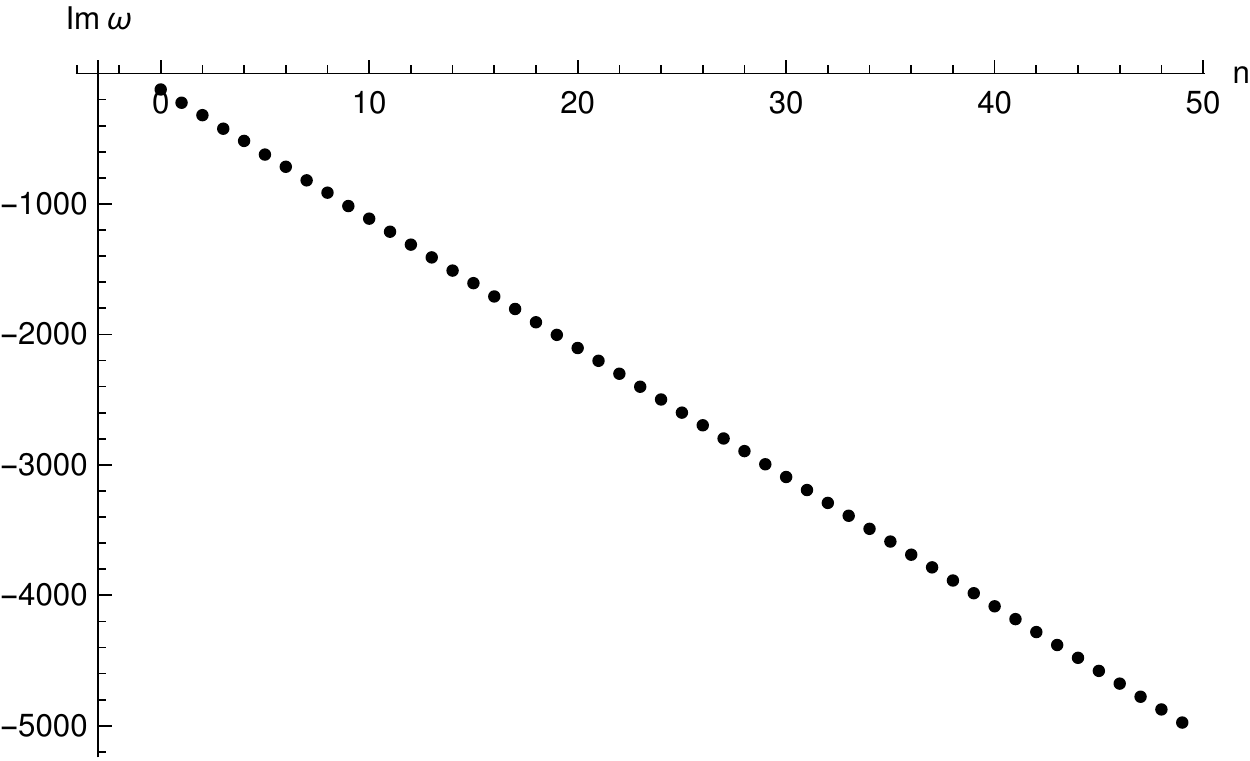}}
  \subfigure[$r_+=50$, $r_-=40$ ($M=4100$, $J=4000$)]{\includegraphics[scale=.4]{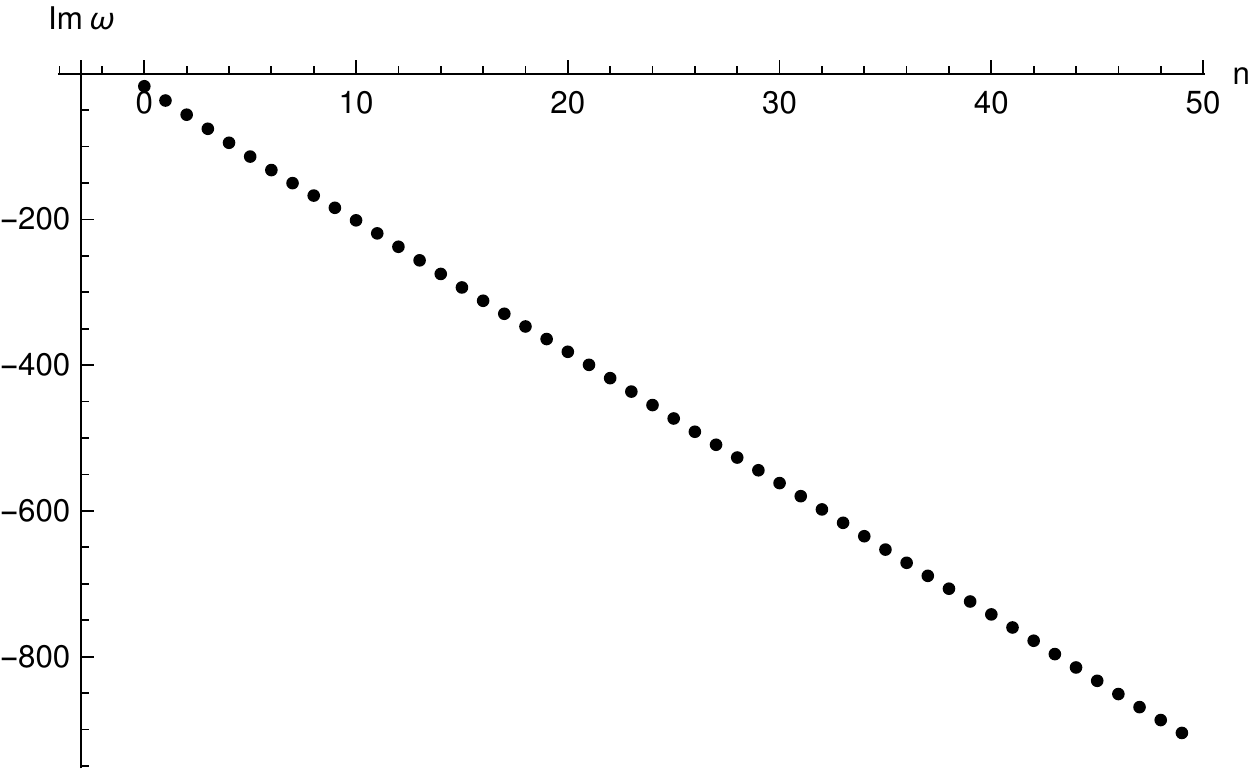}}
  \subfigure[$r_+=100$, $r_-=80$ ($M=16400$, $J=16000$)]{\includegraphics[scale=.4]{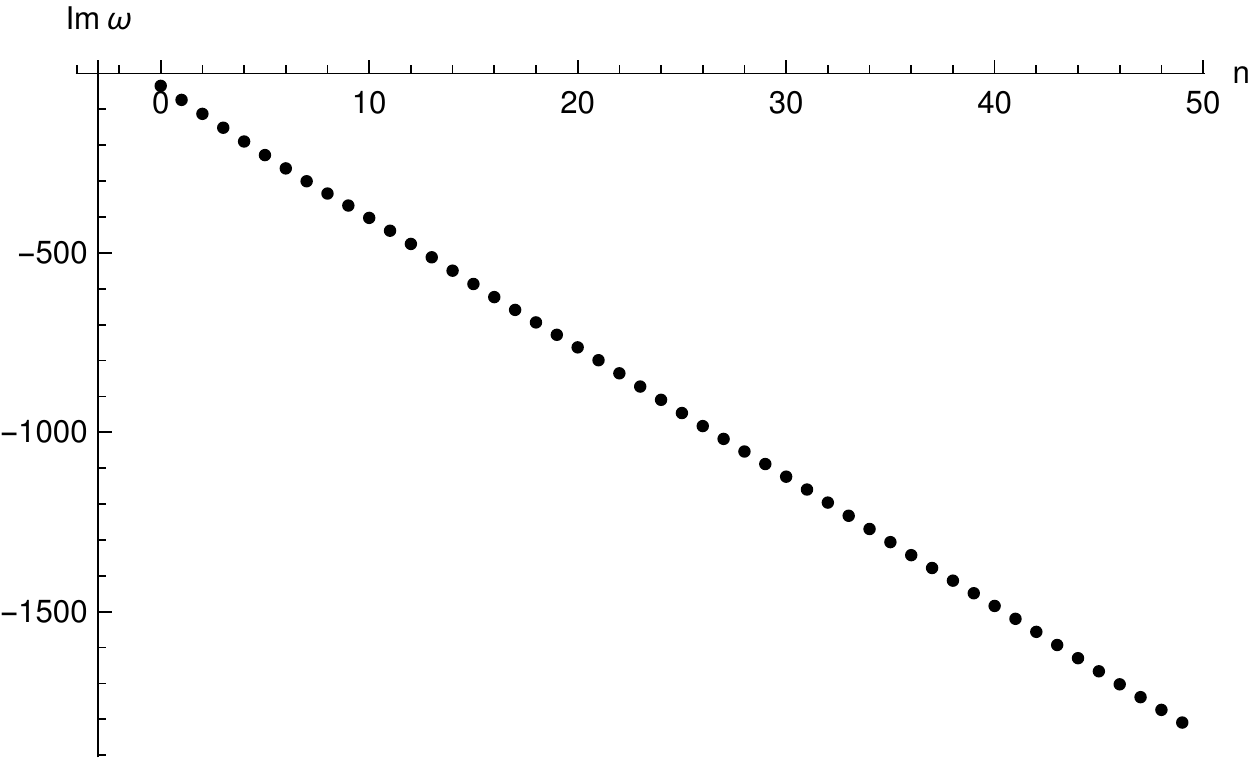}}
   \caption{We show the first fifty QNFs of the Dirac field for different values of the horizons. We take $m=3/4$.}\label{f: dirac n variable}
\end{figure}

\begin{table}[ht]
\centering
\begin{tabular}{cccc}
\hline
$r_+$ & $r_-$ & $\kappa$ & Linear fit  \\  \hline
50 & 10  & 48   &  $ -59.961 - 48.001 n $  \\ 
 100 & 10 & 99  & $ -123.718 - 99.001 n $ \\ 
 50 &  40 & 18  &  $ -21.708 - 18.027 n $ \\ 
 100 &  80 & 36 & $ -43.502 - 36.049 n $ \\ \hline
\end{tabular}
\caption{For the Dirac field we display the linear fits for the QNFs of the first fifty modes of Fig.\ \ref{f: dirac n variable}. We note that this figure shows the dependence of the QNFs on the mode number.}
\label{Tabla 7}
\end{table}

For the Dirac field and four different configurations of the horizons radii, in Fig.\ \ref{f: dirac n variable} we show how the imaginary part of the QNFs depends on the mode number. We notice that the plots ${\mathbb{I}}{\mathrm{m}} (\omega)$ vs $n$ are linear and in Table \ref{Tabla 7} we give the linear fits for the four examples that we display in Fig.\ \ref{f: dirac n variable}. In an analogous way to the Klein-Gordon field, the slope of the straight line is related to the surface gravity (\ref{e: surface gravity}) of the DRBTZ black hole, but the value is not exactly the surface gravity, in contrast to the Dirac field moving in the JT black hole for which the slope of the plot ${\mathbb{I}}{\mathrm{m}} (\omega)$ vs $n$ is the surface gravity (see the expression (\ref{e: dirac qnf uncharged})).

\begin{figure}[ht]
\begin{center}
\includegraphics[scale=.9,clip=true]{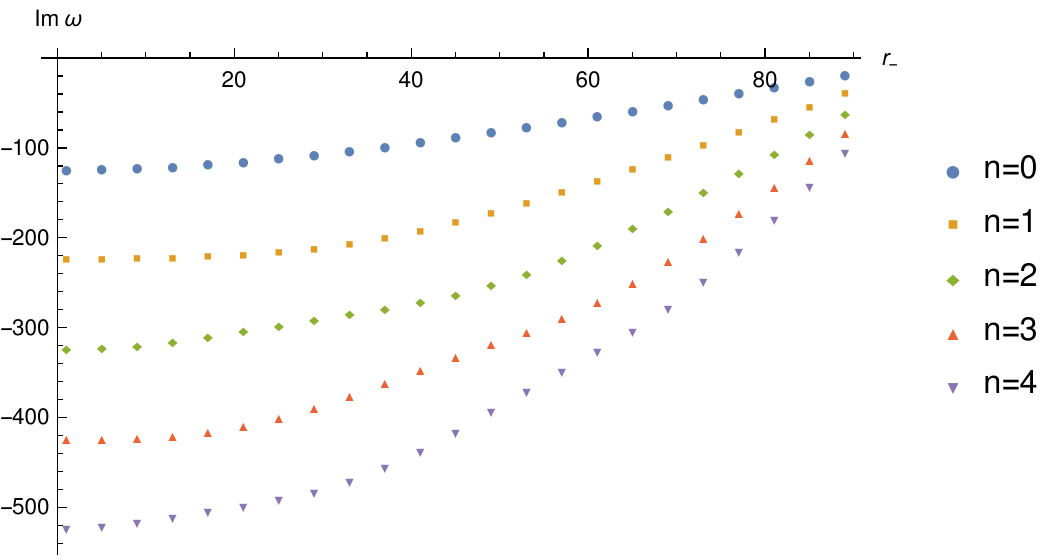}
\caption{We show the dependence on the inner horizon radius of the imaginary part of the QNFs for the Dirac field of mass $m=3/4$. We take $r_+=100$.  \label{f: dirac rmenos variable}} 
\end{center}
\end{figure}

\begin{table}[ht]
\centering
\begin{tabular}{cc}
\hline
$n$ & Quadratic fit  \\  \hline
0 & $ -128.269 + 0.484 r_- + 0.0086 r_-^2 $   \\ 
 1 & $ -226.056- 0.257 r_- + 0.027 r_-^2 $  \\ 
 2 & $ -319.193 - 0.307 r_- + 0.036 r_-^2 $  \\ 
 3 & $ -426.715 + 0.038 r_- + 0.042 r_-^2 $  \\ 
 4 & $ -523.023 - 0.089 r_- + 0.053 r_-^2 $   \\  \hline
\end{tabular}
\caption{For the Dirac field we display the quadratic fits for the QNFs of the first five modes of Fig.\ \ref{f: dirac rmenos variable}. Notice that in this figure we show the dependence of the QNFs on the inner horizon radius.}
\label{Tabla 8}
\end{table}

\begin{table}[ht]
\centering
\begin{tabular}{cccc}
\hline
$n$ & $\omega^a_{JT}$ & $\omega^q_{JT}$ & $\mathcal{E}$  \\  \hline
0 &$-125i$ & $ -128.269i$   & 2.615 \%  \\ 
 1 & $-225 i$& $ -226.056i $ &  0.469 \%  \\ 
 2 &  $-325i $& $  -319.193i $ &  1.787 \% \\ 
 3 & $-425  i$& $ -426.715 i$  &  0.403 \% \\ 
 4 &  $-525 i$ & $  -523.023i$ &  0.376 \% \\  \hline
\end{tabular}
\caption{Based on the quadratic fits of Table \ref{Tabla 8} we give the predicted values for the QNFs of the Dirac field moving in the JT black hole ($J=r_-=0$) whose QNFs are known analytically \cite{Cordero:2012je}. We also show the relative errors.}
\label{Tabla 9}
\end{table}

For the QNFs of the Dirac field moving in the DRBTZ black hole, in Fig.\ \ref{f: dirac rmenos variable} we display the dependence of the imaginary part on the inner horizon radius of the DRBTZ black hole. We notice that for the first five modes the absolute value of the imaginary part decreases as the inner horizon radius increases. As a consequence, for fixed event horizon radius, as the inner horizon increases the decay time increases. Furthermore we observe that for the first five modes the damping  decreases as the inner horizon radius increases.

In Fig.\ \ref{f: dirac rmenos variable} we see that the plots ${\mathbb{I}}{\mathrm{m}} (\omega)$ vs $r_-$ can be described by a quadratic relation and in Table \ref{Tabla 8} we give the quadratic fits for the first five modes of the Fig.\ \ref{f: dirac rmenos variable}. Taking as a basis the quadratic fits of Table \ref{Tabla 8} we can predict the values of the QNFs for the Dirac field moving in the JT black hole. We display the predicted values in Table \ref{Tabla 9} and compare with the  values that produce the analytical expression (\ref{e: dirac qnf uncharged}). In this table, for the first five QNFs we also give their relative errors (defined in the formula (\ref{e: relative error})). We see that the quadratic fits of Table \ref{Tabla 8} produce accurate values for the QNFs of the Dirac field in the JT black hole, but the relative errors for the QNFs of the Dirac field are larger than those we calculate previously for the first five QNFs of the Klein-Gordon field (see Table \ref{Tabla 3}).

\begin{figure}[ht]
\begin{center}
\includegraphics[scale=.9,clip=true]{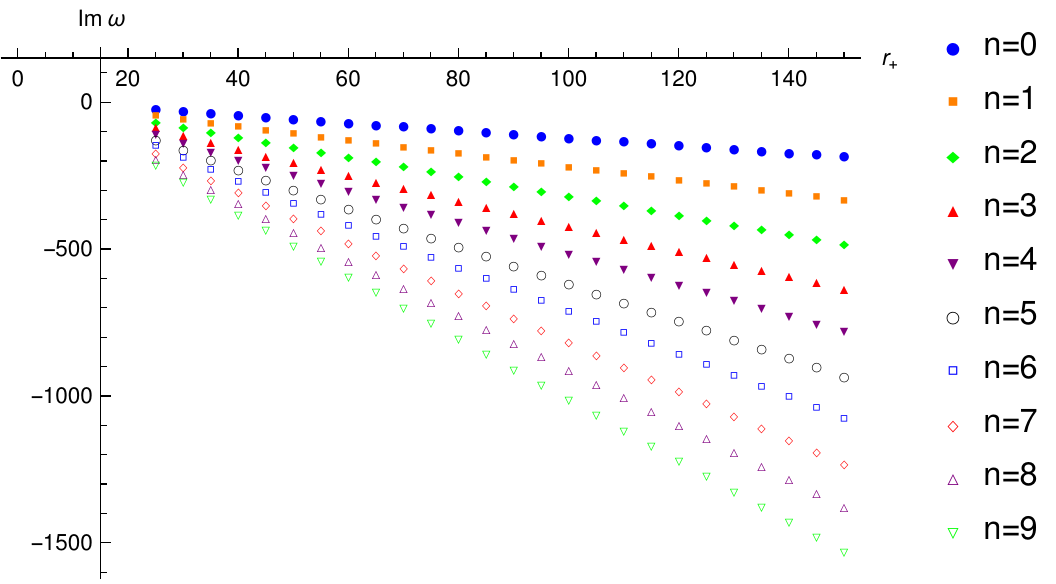}
\caption{We show the dependence on the event horizon radius of the imaginary part of the QNFs for the Dirac field of mass $m=3/4$. We take $r_-=10$.  \label{f: dirac rmas variable}} 
\end{center}
\end{figure}

\begin{table}[ht]
\centering
\begin{tabular}{cc}
\hline
$n$ & Linear fit  \\  \hline
0 &  $ 6.429-1.289 r_+$ \\ 
 1 &  $4.362-2.283 r_+ $ \\ 
 2 &  $ 12.799-3.318 r_+$ \\ 
 3 & $ 12.987-4.350 r_+ $\\ 
 4 & $ 16.877-5.333 r_+ $\\  
 5 & $ 21.227-6.408 r_+$ \\  
 6 & $ 22.292-7.365 r_+ $ \\  
 7 &  $ 28.553-8.453 r_+ $\\  
 8 &  $ 28.138-9.404 r_+ $ \\  
 9 &  $ 35.322-10.491 r_+ $ \\  \hline
\end{tabular}
\caption{For the Dirac field we display the linear fits for the QNFs of the first ten modes of Fig.\ \ref{f: dirac rmas variable}. We observe that this figure shows how the QNFs depend on the event horizon radius.}
\label{Tabla 11}
\end{table}

For the Dirac field propagating in the DRBTZ black hole with inner horizon radius $r_-=10$, in Fig.\ \ref{f: dirac rmas variable} we show how the imaginary part of the QNFs depends on the event horizon radius. We observe that the plots ${\mathbb{I}}{\mathrm{m}} (\omega)$ vs $r_+$ are straight lines in a similar way to the Klein-Gordon field. The linear fits for the graphs of Fig.\ \ref{f: dirac rmas variable} are given in Table \ref{Tabla 11}. We note in this table that the absolute value of the slope for the straight line increases as the mode number increases, in an analogous way to the Klein-Gordon field previously studied. From the analytical expression (\ref{e: dirac qnf uncharged}) for the QNFs of the Dirac field, we deduce that in the JT black hole the slope of similar plots is equal to $-(n+1.25)$ for the same values of the physical parameters. For the first ten modes of the DRBTZ black hole we observe that for the plots ${\mathbb{I}}{\mathrm{m}} (\omega)$ vs $r_+$  the absolute values of their slopes are larger than the absolute values of the slopes for the similar plots of the JT black hole. For the Dirac field,  from Fig.\ \ref{f: dirac rmas variable} we also get that for fixed inner horizon radius, the decay time decreases as the event horizon radius increases. Also for the first ten QNMs the damping increases as the event horizon radius increases.

\begin{figure}[ht]
\begin{center}
\includegraphics[scale=.9,clip=true]{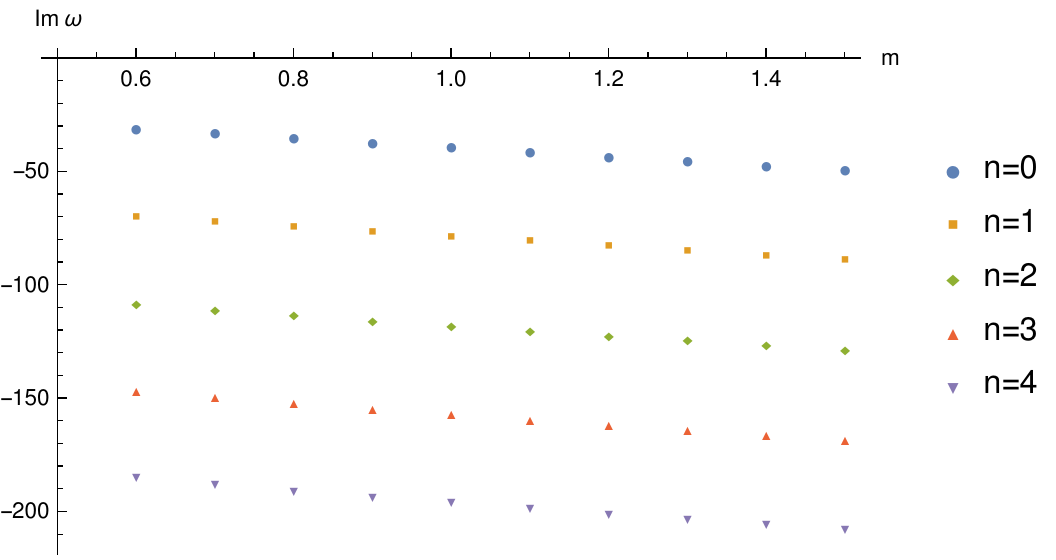}
\caption{We show the dependence on the field mass of the imaginary part of the QNFs for the Dirac field. We take $r_+=100$ and $r_-=80$ ($M=16400$, $J=16000$).  \label{f: dirac mass variable}} 
\end{center}
\end{figure}

\begin{table}[ht]
\centering
\begin{tabular}{cc}
\hline
$n$ & Linear fit  \\  \hline
0 & $ -19.16 - 20.44 m$ \\ 
 1 &  $-57.62 - 21.31 m $ \\ 
 2 &  $ -95.74 - 22.36 m$ \\ 
 3 & $ -133.47 - 23.62 m$ \\ 
 4 & $ -170.63 - 25.23 m$ \\  \hline
\end{tabular}
\caption{For the Dirac field we display the linear fits for the QNFs of the first five modes of Fig.\ \ref{f: dirac mass variable}. Notice that in this figure we show how the QNFs depend on the mass of the field.}
\label{Tabla 12}
\end{table}

In Fig.\ \ref{f: dirac mass variable} we display how the first five QNFs  depend on the mass of the Dirac field. We observe that the graphs ${\mathbb{I}}{\mathrm{m}} (\omega)$ vs $m$ show a linear behavior and in Table \ref{Tabla 12} we give their linear fits. From these linear fits we deduce that the slopes and constant terms of the straight lines depend on the mode number and their absolute values increase as the mode number increases. This behavior is similar to that of the QNFs for the Klein-Gordon field. From Fig.\ \ref{f: dirac mass variable} we obtain that for the Dirac field the decay time depends slightly on its mass, in such a way that the decay time decreases as the mass increases.

For the Dirac field, in each of the Figures \ref{f: dirac rmenos variable}, \ref{f: dirac rmas variable}, \ref{f: dirac mass variable}, we see that  the studied QNFs behave in a similar way when we change the physical parameters of the black hole or of the field.

\section{Discussion}
\label{s: Discussion}

From the analytical expressions (\ref{e: QNF KG UAO}) and (\ref{e: dirac qnf uncharged}) we get that in the JT black hole the QNFs of the Klein-Gordon and Dirac fields are purely imaginary. In an analogous way, our numerical results show that the QNFs of these two fields are purely imaginary in the DRBTZ black hole. Furthermore we find that for these two fields their QNFs satisfy that ${\mathbb{I}}{\mathrm{m}} (\omega) < 0$ and as a consequence their QNMs are stable in the JT and DRBTZ black holes. We notice that for the Dirac field propagating in the DRBTZ black hole its QNFs are not previously calculated and our numerical results show that the two potentials (\ref{e: potenciales efectivos Dirac}) have the same spectrum of QNFs.

For the Klein-Gordon and Dirac fields, from our numerical results we notice that in the DRBTZ black hole, as the inner horizon radius increases the damping of the QNMs decreases, that is, the decay time increases as the inner horizon radius increases. For fixed inner horizon radius, as the event horizon radius increases we find that the damping increases. As a consequence, for both fields, for a given mode number, and for fixed event horizon radius, as we change the inner horizon radius, the more damped QNM occurs for the JT black hole. Furthermore, taking as a basis the linear fits of Tables \ref{Tabla 5} and \ref{Tabla 11}, for the Klein-Gordon and Dirac fields we find that for $r_+  >> r_-$ the predicted values   for the QNFs of the DRBTZ black hole tend to the QNFs of the JT black hole of the same event horizon radius. 

For the Klein-Gordon and Dirac fields moving in the DRBTZ black hole, we notice that the effective potentials (\ref{e: efective potential KG}) and (\ref{e: potenciales efectivos Dirac}) are different, nevertheless our numerical results show that the behavior of the QNFs for both fields is similar when we change the physical parameters. We notice that this conclusion follows from the numerical results. Considering the mathematical form of the effective potentials (\ref{e: efective potential KG}) and (\ref{e: potenciales efectivos Dirac}) is not straightforward to deduce that their spectra of QNFs behave in a similar form.

\begin{figure}[ht]
\begin{center}
\includegraphics[scale=.8,clip=true]{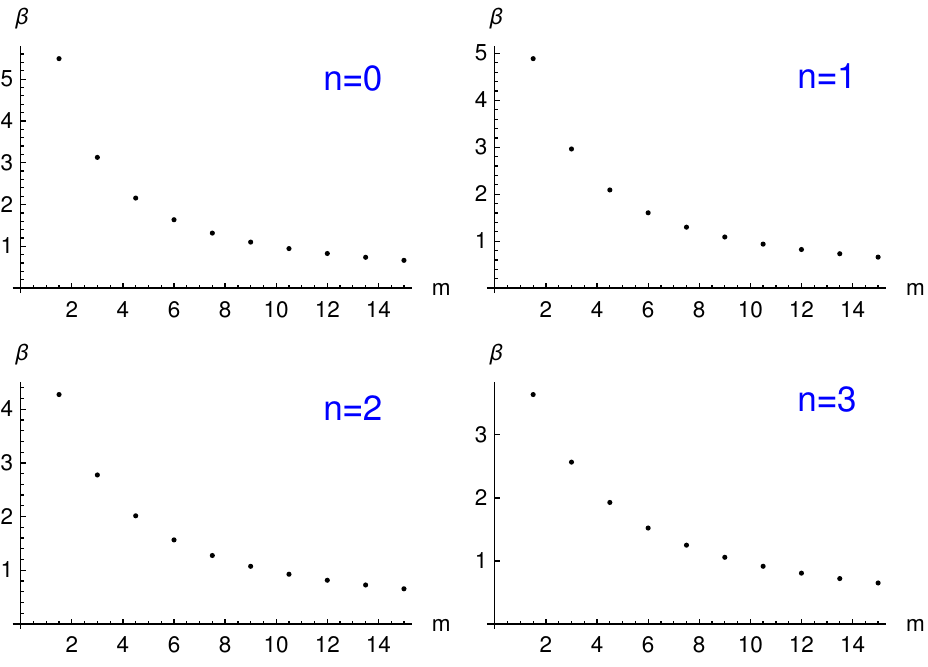}
\caption{For the DRBTZ black hole with $r_+=100$, $r_-=80$ ($M=16400$, $J=16000$), and for the first four QNMs, we show the absolute value of the difference of the values for the QNFs of the Dirac and Klein-Gordon fields  as the masses of the fields increase.   \label{f: diferencias de masa}}
\end{center}
\end{figure}

Furthermore, from the analytical expressions (\ref{e: QNF KG UAO}) and (\ref{e: dirac qnf uncharged}) for the QNFs of the JT black hole, we notice that for large mass of the field both expressions behave in the form
\begin{equation}
 \omega = -i \kappa \left( n + \frac{1}{2} + m \right).
\end{equation} 
That is, for the JT black hole, the QNFs of the Klein-Gordon and Dirac fields are almost the same for large values of the field mass. For the DRBTZ black hole our numerical results for the first four modes are shown in Fig.\ \ref{f: diferencias de masa}, where we plot the quantity $\beta$ defined by
\begin{equation}
 \beta = \lvert \omega_D - \omega_{KG}  \lvert
\end{equation} 
versus the mass of the field. In the previous expressions $\omega_D$ denotes the QNFs of the Dirac field and $\omega_{KG} $ stands for the QNFs of the Klein-Gordon field.

In Fig.\ \ref{f: diferencias de masa} we observe that the quantity $\beta$ decreases as the masses of the fields increase, thus, the difference between the values the QNFs of the Klein-Gordon and Dirac fields decreases as the masses of fields increase. Thus, the QNFs of the DRBTZ black hole behave in a similar way to the QNFs of the JT black hole as the mass of the test field increases.

In the JT black hole, from the analytical expressions (\ref{e: QNF KG UAO}) and (\ref{e: dirac qnf uncharged}) we see that the QNFs of the Klein-Gordon and Dirac fields are different for small values of the mass of the field. In Table \ref{Tabla 13} we give the QNFs for the first ten modes of the Klein-Gordon and Dirac fields with mass equal to $3/4$ propagating in the DRBTZ black hole with $r_+=100$, $r_-=80$, and we see that the QNFs of these fields are different, in contrast to some spacetimes for which the QNFs of the Klein-Gordon and Dirac fields are similar \cite{Kokkotas:1999bd}--\cite{Berti:2009kk}. Thus, for small values of the mass of the field, in the DRBTZ black hole the QNFs of these two fields are different and this behavior is similar to that of the QNFs for the BTZ three-dimensional black hole \cite{Birmingham:2001pj}, \cite{Cardoso:2001hn},  from which the DRBTZ black hole is a dimensional reduction. Nevertheless, from the data shown in Table \ref{Tabla 13}, we notice that for both fields, as the mode number increases the QNFs of the DRBTZ black hole approach those of the JT black hole with the same surface gravity.

\begin{table}[ht]
\centering
\begin{tabular}{ccccc}
\hline
$n$ & $\omega_{KG}$  DRBTZ  & $\omega_{KG}$ JT & $\omega_{D}$ DRBTZ  & $\omega_{D}$ JT  \\  \hline
0& -42.980 $i$& -50.450 $i$& -34.499 $i$& -45.000$i$\\
1& -80.834 $i$& -86.450 $i$& -73.613 $i$& -81.000$i$\\
2& -118.571 $i$& -122.450 $i$& -112.521 $i$& -117.000$i$\\
3& -156.151  $i$& -158.450 $i$& -151.205 $i$& -153.000$i$\\
4& -193.513 $i$& -194.450 $i$& -189.577 $i$& -189.000$i$\\
5& -230.580 $i$& -230.450 $i$& -227.460 $i$& -225.000$i$\\
6& -267.270 $i$& -266.450 $i$& -264.538 $i$& -261.000$i$\\
7& -303.524  $i$& -302.450 $i$& -300.329$i$& -297.000$i$\\
8& -339.355 $i$& -338.450 $i$& -334.548 $i$& -333.000$i$\\
9& -374.891 $i$& -374.450 $i$& -368.045$i$& -369.000$i$\\
\hline
\end{tabular}
\caption{In the second and fourth columns we give the QNFs of the Klein-Gordon and Dirac fields with mass $m=3/4$  propagating in the DRBTZ black hole with $r_+=100$ and $r_-=80$ (that is, $\kappa=36$, $M=16400$, and $J=16000$). In the third and fifth columns we show the QNFs of the Klein-Gordon and Dirac fields with mass $m=3/4$  propagating in the JT black hole with surface gravity $\kappa=36$, that is, with radius of the horizon equal to $r_+ = 36$.}
\label{Tabla 13}
\end{table}

In addition, we note the following. For the JT black hole we find that for the massive generalized Klein-Gordon field its QNFs take the form (see the formula (3.13) of Ref.\ \cite{Cadoni:2021qfn} with $\alpha = 1$ and $L=1$)
\begin{equation}
 \omega_{GKG} = -2 i r_+ \left( n + \frac{1}{2} + \frac{1}{2} \sqrt{1 + m^ 2} \right),
\end{equation} 
and therefore the spacing between consecutive QNFs is equal to
\begin{equation}
 \Delta \omega_{GKG} = - i 2 r_+ .
\end{equation} 
From Eq.\ (\ref{e: QNF KG UAO}) we get that the spacing for the massive minimally coupled Klein-Gordon field is equal to
\begin{equation}
 \Delta \omega_{KG} = - i r_+,
\end{equation} 
which is different from the spacing for the generalized Klein-Gordon field studied in Refs.\ \cite{Bhattacharjee:2020nul}, \cite{Cadoni:2021qfn}.

For the DRBTZ black hole, in Ref.\ \cite{Bhattacharjee:2020nul} they find that the QNFs of the generalized massive Klein-Gordon field are of two types, the left and the right QNFs given by (see the expressions (4.19a) and (4.19b) of Ref.\ \cite{Bhattacharjee:2020nul})
\begin{equation}
 \omega_L = -i (r_+ - r_-)(\tilde{\Delta} + 2 n), \qquad \qquad \omega_R = -i (r_+ + r_-)(\tilde{\Delta} + 2 n),
\end{equation} 
where $m^2=\tilde{\Delta} (\tilde{\Delta} -2)$. As a consequence, we get that the spacings between consecutive QNFs are
\begin{equation} \label{e: spacings DRBTZ GKG}
 \Delta \omega_L = -i 2 (r_+ - r_-), \qquad \qquad \Delta \omega_R = -i 2 (r_+ + r_-).
\end{equation} 
Our numerical results point out that the spacing of consecutive QNFs of the minimally coupled Klein-Gordon field is proportional to the surface gravity (\ref{e: surface gravity}) as the mode number increases and therefore different from the spacings (\ref{e: spacings DRBTZ GKG}) for the generalized massive Klein-Gordon field. Thus, the QNFs of the minimally coupled Klein-Gordon field and of the generalized Klein-Gordon field have some different characteristics in the 2D DRBTZ black hole. 

Since the spacing of the QNFs for the minimally coupled Klein-Gordon field in the static (rotating) BTZ black hole is identical \cite{Birmingham:2001pj}, \cite{Cardoso:2001hn} to the spacing of the QNFs for the generalized Klein-Gordon field propagating in the 2D JT (DRBTZ) black hole, we can deduce that, for the minimally coupled Klein-Gordon field, the BTZ black hole and the dimensional reduction proposed by Achucarro-Ortiz react in a different way.  

Finally, for the 2D DRBTZ black hole, we should investigate the relevance of our numerical results to the discussion of the AdS/CFT correspondence, to the calculation of the microscopic structure of the black hole \cite{Cadoni:2021qfn}, and to the determination of the spectrum of the event horizon \cite{Hod:1998vk}, \cite{Maggiore:2007nq}, although it is likely that we should extend our computations to the asymptotic regime.

\section{Acknowledgments}

This work was supported by CONACYT M\'exico, SNI M\'exico, EDI IPN, COFAA IPN, and Research Project IPN SIP 20221379.

\begin{appendix}

\section{Asymptotic iteration method}
\label{a: aim method}

An useful method to find the eigenvalues of linear second order differential equations is the asymptotic iteration method \cite{Ciftci Hall and Saad 2003}--\cite{Cho:2011sf}. This method works with linear second order differential equations of the form 
\begin{equation}\label{e: aim equation}
y''= \lambda _0 (x)y'+s_0(x)y,
\end{equation}
where the prime denotes differentiation with respect to the independent variable. Taking advantage of the symmetrical structure of Eq.\ (\ref{e: aim equation}) we find that its first derivative takes a similar mathematical form 
\begin{equation}\label{e: aim equation 2}
y'''= \lambda _1 (x)y'+s_1(x)y,
\end{equation}
with
\begin{equation}\label{e:aim equation 3}
\lambda _1 =\lambda _0' + s_0 + \lambda _0^2, \quad \textrm{and} \quad s_1=s_0'+s_0 \lambda _0.
\end{equation}
In an analogous way we find that the $(n+2)$-th derivative of the function $y$ is equal to \cite{Ciftci Hall and Saad 2003}--\cite{Cho:2011sf}
\begin{equation}\label{e: aim equation 7}
y^{(n+2)}= \lambda _{n} (x)y'+s_{n}(x)y,
\end{equation}
where
\begin{equation}\label{e: aim lambda s n}
\lambda _n =\lambda _{n-1}' + s_{n-1} + \lambda _0 \lambda _{n-1}, \quad \textrm{and} \quad s_n=s_{n-1}'+s_0 \lambda _{n-1}.
\end{equation}
The asymptotic aspect of the method imposes that for sufficiently large $n$ the functions $\lambda_n$ and $s_n$ satisfy \cite{Ciftci Hall and Saad 2003}--\cite{Cho:2011sf}
\begin{equation}\label{e: aim condition}
\frac{s_n}{\lambda _n} =\frac{s_{n-1}}{\lambda _{n-1}} = \alpha
\end{equation}
or in an equivalent form
\begin{equation} \label{e: aim condition 2}
 s_n \lambda _{n-1} - s_{n-1} \lambda _n = 0.
\end{equation} 
Solving this equation, usually known as discretization condition, we can find the QNFs of the black hole under study \cite{Cho:2009cj}, \cite{Cho:2011sf}. 

Nevertheless, the computation of the recurrence relations (\ref{e: aim lambda s n}) requires many resources \cite{Ciftci Hall and Saad 2003}--\cite{Cho:2011sf}. Owing this fact, in Ref.\ \cite{Cho:2009cj} is proposed an improved version of the AIM. In this version of the method we expand the functions $\lambda_n$ and $s_n$ around a convenient point $\xi$, that is,
\begin{equation}\label{e: serie lambda s aim}
\lambda _n (\xi )= \sum ^\infty _{i=0} c^i _n (x- \xi )^i, \qquad \qquad 
s _n (\xi )= \sum ^\infty _{i=0} d^i _n (x- \xi )^i,
\end{equation}
to find that the recurrence relations (\ref{e: aim lambda s n}) imply that the coefficients $c^i _n$ and $d^i _n$ satisfy 
\begin{align}\label{eq:coefcaim}
c^i _n &= (i+1)c^{i+1} _{n-1} + d^i _{n-1} + \sum^i _{k=0} c^k _0 c^{i-k} _{n-1}, \nonumber \\
d^i _n &= (i+1) d^{i+1} _{n-1} + \sum ^i _{k=0} d^k _0 c^{i-k} _{n-1}.
\end{align}
Furthermore, we get that the discretization condition (\ref{e: aim condition 2}) takes the form
\begin{equation}\label{e: final AIM}
d^0 _n c^0 _{n-1} - d^0 _{n-1} c^0 _n =0.
\end{equation}
Solving this equation for different values of $n$, we numerically obtain the QNFs of the studied black hole \cite{Cho:2009cj}, \cite{Cho:2011sf}. In this work we use this improved formulation of the AIM to find the QNFs of the Klein-Gordon and Dirac fields propagating in the DRBTZ black hole.

\section{Horowitz-Hubeny method for the Dirac field}
\label{a: HH Dirac}

For the Klein-Gordon field moving in the DRBTZ black hole,  in Ref.\ \cite{Cordero:2012je} are calculated its QNFs taking as a basis the Horowitz-Hubeny method. Nevertheless a restriction on the values of the horizons radii was imposed, as we previously commented. For the Dirac field we can use the Horowitz-Hubeny method to compute its QNFs in the DRBTZ black hole. In what follows we describe the steps to transform the radial equations (\ref{e: schrodinger dirac}) into a convenient form to use the Horowitz-Hubeny method \cite{Horowitz:1999jd}.

First, to fulfill the boundary condition near the event horizon, we propose that the solutions of Eqs.\ (\ref{e: schrodinger dirac}) take the form
\begin{equation}\label{e: R new}
R_s(r)=e^{-i(\omega \pm \frac{i \kappa}{2})r_*} U_s(r),
\end{equation}
to find that the functions $U_s$ must satisfy the differential equations 
\begin{equation}\label{e: radial HH Dirac}
f\frac{d^2U_s}{dr^2}+\Big(\frac{df}{dr}-2i\omega\pm\kappa\Big)\frac{dU_s}{dr}+\frac{1}{f}\Big(\frac{\kappa^2}{4}\mp i\omega\kappa-V_s \Big)U_s=0.
\end{equation}
To have an independent variable that changes in a finite interval, following to Horowitz-Hubeny \cite{Horowitz:1999jd}, we make the change of variable
\begin{equation}\label{e: x HH}
x=\frac{1}{r},
\end{equation}
to get that the radial equations (\ref{e: radial HH Dirac}) transform into
\begin{equation}\label{e: HH equation Dirac}
S_s(x)\frac{d^2U_s}{dx^2}+\frac{t_s(x)}{x-x_+}\frac{dU_s}{dx}+ \frac{u_s(x)}{(x-x_+)^2}U_s=0,
\end{equation}
where the functions $t_s(x)$, $u_s(x)$, and $S_s(x)$  are equal to 
\begin{align} \label{e: t u s x dirac}
 t_s(x)&= 2x(x^2-x_+^2)(x^2-x_-^2)^2(x + x_+) + 2x^5(x + x_+)(x^2 - x_-^2) \nonumber \\
   &-2 x x_+^2 x_-^2 (x + x_+)(x^2 - x_-^2) + 2i\omega x^2(x+x_+)(x^2-x_-^2)x_+^2 x_-^2 \nonumber \\
   &\mp x^2x_+(x_-^2-x_+^2)(x+x_+)(x^2-x_-^2), \nonumber \\
 u_s(x)&= \frac{1}{4}x^2 x_+^2(x_+^2-x_-^2)^2 \mp i \omega x^2x_+^3x_-^2(x_-^2-x_+^2) \nonumber \\
   &\pm i \omega xx_+^2x_-^2(x_+^2x_-^2-x^4) + \frac{5}{4}x^8 - \frac{3}{2} x^6(x_+^2+x_-^2) + \frac{5}{2}x^4x_+^2x_-^2 \nonumber \\
   &- \frac{1}{2} x^2 x_+^2 x_-^2 (x_+^2 + x_-^2) + \frac{1}{4} x_+^4x_-^4-m^2(x^2-x_+^2)(x^2-x_-^2)x_+^2x_-^2,\nonumber \\
   S_s(x) &=  x^2(x+x_+)^2(x^2-x_-^2)^2.
\end{align}
In the previous equations we define $x_+= 1/r_+$ and $x_-=1/r_-$.
   
As in Ref.\ \cite{Horowitz:1999jd} we propose that the solutions to Eqs.\ (\ref{e: HH equation Dirac}) take the form
\begin{equation}\label{e: Us serie}
U_s=(x-x_+)^{\nu_s} \sum_{k=0}^{\infty} a_{k,s} (\omega )(x-x_+)^k,
\end{equation} 
but to fulfill the boundary condition of the QNMs near the event horizon we must take $\nu_s=0$ \cite{Horowitz:1999jd}. Substituting these simplified forms of the functions $U_s$  into Eq.\ (\ref{e: HH equation Dirac}), we find that the coefficients $a_{k,s}$ must satisfy the recurrence relations
\begin{equation}\label{e: ak(w) coefficients}
a_{k,s}=-\frac{1}{k(k-1)S_{0,s}+k t_{0,s}} \sum_{n=0}^{k-1} a_{n,s} (n(n-1)S_{k-n,s}+n t_{k-n,s}+u_{k-n,s}),
\end{equation}
where $ S_{0,s}= S_s (x_+)$, $ t_{0,s} = t_s(x_+)$, the coefficients $u_{k,s}$ are given by
\begin{equation}\label{e: u serie}
u_s(x)= \sum_{k=0}^{\infty} u_{k,s} (x-x_+)^k,
\end{equation}
and similar definitions are valid for the coefficients $S_{k,s}$ and $t_{k,s}$.

At the asymptotic region, the boundary condition of the QNMs  imposes that as $x \to 0$ ($r \to \infty$) the functions $U_s$ go to zero, that is,
 \begin{equation}\label{e: final HH}
U_s(x=0)=\sum_{k=0}^{\infty} a_{k,s} (\omega)(-x_+)^k=0
\end{equation}  
and the QNFs of the Dirac field can be calculated by finding the roots of the previous equation  when we replace the infinite sum by a finite sum  \cite{Horowitz:1999jd}. 

As for the Klein-Gordon field moving in the DRBTZ black hole, the Horowitz-Hubeny method allows us to calculate the QNFs of the Dirac field when $r_+ >  2r_-$. The reason is that the series (\ref{e: Us serie}) has  a radius of convergence as large as the distance to the nearest singular point. Since the radius of convergence must include the point $x=0$, the distance from the expansion point at $x_+$ to the other singularity at $x_-$ must be larger than the distance between $x_+$ and 0, that is, $x_--x_+ > x_+$, or equivalently $r_+ >  2r_-$.


\end{appendix}

\end{document}